%% file: mix_spin_arxiv_v3.tex
\documentclass[12pt]{article}
\pdfoutput=1
\usepackage[bookmarks=true,hyperfigures=true]{hyperref}
\usepackage{amsfonts}
\usepackage[dvips]{graphicx}
\usepackage{amsmath,amssymb,mathtools,slashed}
\usepackage[nosort]{cite}
\usepackage[bulletsep]{collref}
\usepackage{color}

\usepackage[margin=2.5cm]{geometry}

\newcommand{\be}{\begin{eqnarray}}
\newcommand{\ee}{\end{eqnarray}}
\newcommand{\la}{\lambda}
\newcommand{\NN}{\mathcal{N}}
\newcommand{\CC}{\mathcal{C}}

\newcommand{\GG}{\mathcal{G}}
\newcommand{\OO}{\mathcal{O}}
\newcommand{\s}{\sigma}
\newcommand{\half}{\sfrac{1}{2}}
\newcommand{\al}{\alpha}
\newcommand{\p}{\partial}
\newcommand{\tQ}{\widetilde Q}
\newcommand{\tS}{\widetilde S}
\newcommand{\dda}{{\dot{\alpha}}}

\newcommand{\ve}{\varepsilon}
\newcommand{\tM}{{\tilde M}}
\newcommand{\tN}{{\tilde N}}
\newcommand{\tL}{{\tilde L}}
\newcommand{\slk}{\slashed k}
\newcommand{\slpsi}{\slashed\psi}
\newcommand{\eps}{\varepsilon}

\newcommand{\nn}{\nonumber}

\newcommand{\op}[1]{\mathcal{#1}}
\newcommand{\transpose}{^\mathsf{T}}
\newcommand{\alphap}{\alpha^\prime}

\numberwithin{equation}{section}

\providecommand{\hypersetup}[1]{}
\providecommand{\hyperref}[2][]{#2}

\providecommand{\pdfbookmark}[3][]{}
\hypersetup{plainpages=false}
\hypersetup{pdfpagemode=UseOutlines}
\hypersetup{bookmarksnumbered=true}
\hypersetup{bookmarksopen=true}
\hypersetup{pdfstartview=FitH}
\hypersetup{colorlinks=false}
\hypersetup{citebordercolor={.6 .9 .6}}
\hypersetup{urlbordercolor={.7 .8 1}}
\hypersetup{linkbordercolor={1 .7 .7}}
\hypersetup{pdfborder={0 0 .5}}

\makeatletter
\let\@keywords\@empty
\let\@subject\@empty
\providecommand{\keywords}[1]{\gdef\@keywords{#1}}
\providecommand{\subject}[1]{\gdef\@subject{#1}}
\def\thetitle{\@title}
\def\theauthor{\@author}
\def\thesubject{\@subject}
\def\thedate{\@date}
\def\thekeywords{\@keywords}
\makeatother
\AtBeginDocument{
\hypersetup{pdftitle={\thetitle}}%
\hypersetup{pdfauthor={\theauthor}}%
\hypersetup{pdfsubject={\thesubject}}%
\hypersetup{pdfkeywords={\thekeywords}}%
}

\makeatletter
\def\mr@ignsp#1 {\ifx\:#1\@empty\else #1\expandafter\mr@ignsp\fi}%
\newcommand{\multiref}[1]{\begingroup
\xdef\mr@no@sparg{\expandafter\mr@ignsp#1 \: }%
\def\mr@comma{}%
\@for\mr@refs:=\mr@no@sparg\do{\mr@comma\def\mr@comma{,}\ref{\mr@refs}}%
\endgroup}
\makeatother

\renewcommand{\eqref}[1]{(\multiref{#1})}

\newcommand{\namedref}[2]{\hyperref[#2]{#1~\ref{#2}}}
\newcommand{\secref}[1]{\namedref{Section}{#1}}
\newcommand{\appref}[1]{\namedref{Appendix}{#1}}

\newcommand{\figref}[1]{\namedref{Figure}{#1}}

\let\oldbfseries=\bfseries
\let\oldmdseries=\mdseries
\let\oldnormalfont=\normalfont
\renewcommand{\bfseries}{\oldbfseries\boldmath}
\renewcommand{\mdseries}{\oldmdseries\unboldmath}
\renewcommand{\normalfont}{\oldnormalfont\unboldmath}

\newcommand{\earel}[1]{\mathrel{}&\hspace{-2\arraycolsep}#1\hspace{-2\arraycolsep}&\mathrel{}}
\newcommand{\eq}{\earel{=}}
\newcommand{\eqv}{\earel{\equiv}}

\allowdisplaybreaks[4]

\DeclareMathSymbol{\Gamma}{\mathalpha}{letters}{"00}
\DeclareMathSymbol{\Delta}{\mathalpha}{letters}{"01}
\DeclareMathSymbol{\Theta}{\mathalpha}{letters}{"02}
\DeclareMathSymbol{\Lambda}{\mathalpha}{letters}{"03}
\DeclareMathSymbol{\Xi}{\mathalpha}{letters}{"04}
\DeclareMathSymbol{\Pi}{\mathalpha}{letters}{"05}
\DeclareMathSymbol{\Sigma}{\mathalpha}{letters}{"06}
\DeclareMathSymbol{\Upsilon}{\mathalpha}{letters}{"07}
\DeclareMathSymbol{\Phi}{\mathalpha}{letters}{"08}
\DeclareMathSymbol{\Psi}{\mathalpha}{letters}{"09}
\DeclareMathSymbol{\Omega}{\mathalpha}{letters}{"0A}

\newcommand{\brk}[1]{(#1)}

\newcommand{\bigbrk}[1]{\bigl(#1\bigr)}
\newcommand{\biggbrk}[1]{\biggl(#1\biggr)}

\newcommand{\vev}[1]{\langle#1\rangle}
\newcommand{\bigvev}[1]{\bigl\langle#1\bigr\rangle}

\newcommand{\bigacomm}[2]{\big\{#1,#2\big\}}

\newcommand{\grp}[1]{\mathrm{#1}}

\newcommand{\indup}[1]{_{\mathrm{#1}}}

\newcommand{\supup}[1]{^{\mathrm{#1}}}

\newcommand{\sfrac}[2]{{\textstyle\frac{#1}{#2}}}

\newcommand{\gc}{g\indup{c}}
\newcommand{\tr}{\mathop{\mathrm{Tr}}}
\newcommand{\Tr}{\mathop{\mathrm{Tr}}}

\newcommand{\superN}{\mathcal{N}}
\newcommand{\sym}{$\superN=\nolinebreak4$ SYM}

\newcommand{\ap}{\alpha'}

\newcommand{\sapt}{\sfrac{\alpha'}{2}}

\newcommand{\stap}{\sfrac{2}{\alpha'}}

\title{Three-point correlators from string amplitudes: \\Mixing and Regge spins}
\author{ Joseph A. Minahan and Raul Pereira}
\subject{Gauge Theory, String Theory, AdS/CFT}

\begin{document}

\pdfbookmark[1]{Title Page}{title}

\thispagestyle{empty}
\setcounter{page}{0}
\begin{flushright}\footnotesize
\texttt{UUITP-12/14}
\end{flushright}

\vspace{0.5cm}

\begin{center}
{\Large\textbf{Three-point correlators from string amplitudes: \\Mixing and Regge spins} \par}

\vspace{12mm}

\small

\textsc{ Joseph~A.~Minahan${}$ and Raul Pereira${}$}

\bigskip

\medskip

\textit{Department of Physics and Astronomy, Uppsala University\\
Box 516, SE-751 20 Uppsala, Sweden}

\bigskip

\texttt{ joseph.minahan \& raul.pereira@physics.uu.se}

\normalsize

\vspace{15mm}

\textbf{Abstract}

\vspace{5mm}

\begin{minipage}{12cm}

This paper has two parts.  We first compute the leading contribution to the strong-coupling mixing between the Konishi operator and a double-trace operator composed of chiral primaries by using flat-space vertex operators for the string-duals of the operators.  We then compute the three-point functions for protected or unprotected scalar operators with higher spin operators on the leading Regge trajectory.  Here we see that the nontrivial spatial structures required by conformal invariance arise naturally from the form of the polarization tensors in the vertex operators. We find agreement with recent results extracted from Mellin amplitudes for four-point functions, as well as with earlier supergravity calculations.  We also obtain some new results for other combinations of operators.

\end{minipage}

\end{center}

\newpage

\renewcommand{\thefootnote}{\arabic{footnote}}

\setcounter{tocdepth}{3}
\hrule height 0.75pt
\pdfbookmark[1]{\contentsname}{contents}
{\small\tableofcontents}
\vspace{0.8cm}
\hrule height 0.75pt
\vspace{1cm}

\section{Introduction}

We can now compute  the dimensions of single-trace local operators in planar $\NN=4$ super Yang-Mills (SYM),  at least in principle, because of its underlying integrability \cite{Beisert:2010jr}. However, to fully solve the theory it is also necessary to know the three-point functions between local operators. There are now a number of important results on this subject at both  weak   \cite{Roiban:2004va,Alday:2005nd,Alday:2005kq,Escobedo:2010xs,Escobedo:2011xw,Gromov:2011jh,Bissi:2011ha,Kostov:2012yq,Kazakov:2012ar, Gromov:2012uv,Foda:2013nua, Vieira:2013wya,Caetano:2014gwa, Jiang:2014mja} and strong coupling  \cite{Zarembo:2010rr,Costa:2010rz, Janik:2010gc,Buchbinder:2011jr,Bissi:2011dc,Georgiou:2011qk, Janik:2011bd,Kazama:2011cp,Kazama:2012is,Kazama:2013qsa,Alday:2013cwa, Bajnok:2014sza}.

In this paper we continue our study of three-point 
correlators 
 in $\NN=4$ SYM at strong coupling by using flat-space 
vertex operators  
to find the relevant couplings \cite{Minahan:2012fh,Bargheer:2013faa}.  Here we  investigate two issues.  The first involves the appearance of poles in extremal three-point amplitudes.  This leads to a mixing between single- and double-trace operators which we explicitly compute for a Konishi operator mixing with a double-trace operator composed of two chiral primaries.

The second issue we consider is the three-point functions for scalar operators with a higher spin operator.  Using  results of Schlotterer for flat-space massive superstring amplitudes \cite{Schlotterer:2010kk}, we can straightforwardly compute the three-point functions and compare with recent results in \cite{Costa:2012cb}, where the authors study these correlators by considering Regge amplitudes of four scalar operators.  In the limit of large dimensions we find agreement with their results.

In a conformal field theory, the correlator of three local 
scalar 
operators must have the form
\be\label{3corr1}
\bigvev{\OO_{\Delta_1}(x^\mu_1)\OO_{\Delta_2}(x^\mu_2)\OO_{\Delta_3}(x^\mu_3)}\eq\frac{
\mathcal{C}_{123} }{|x_{12}|^{\Delta_1+\Delta_2-\Delta_3}|x_{23}|^{\Delta_2+\Delta_3-\Delta_1}|x_{31}|^{\Delta_3+\Delta_1-\Delta_2}}\,,
\ee
where $\Delta_i$ are the operator dimensions.  Using Witten diagrams \cite{Witten:1998qj}, one finds that the structure constant $\mathcal{C}_{123} $ in a four-dimensional CFT has the form \cite{Freedman:1998tz}
\be\label{3corr2}
\mathcal{C}_{123}=\frac{\sqrt{(\Delta_1-1)(\Delta_2-1)(\Delta_3-1)}}{2^{5/2}\pi}\,\frac{\Gamma(\al_1)\Gamma(\al_2)\Gamma(\al_3)\Gamma(\Sigma-2)}{\Gamma(\Delta_1)\Gamma(\Delta_2)\Gamma(\Delta_3)}\,\GG_{123}\,,
\ee
where
\be
\label{eq:alphasigma}
\Sigma=\half(\Delta_1+\Delta_2+\Delta_3) \,, \qquad\alpha_i = \Sigma - \Delta_i \,.
\ee
The coupling $\GG_{123}$ is given by
\be\label{coupling}
\GG_{123}=
\frac{8\pi}{\gc^2\ap}\,\vev{V_{k_1}V_{k_2}V_{k_3}}\,\vev{\psi_{J_1}\psi_{J_2}\psi_{J_3}}\,,
\ee
where 
$\vev{\psi_{J_1}\psi_{J_2}\psi_{J_3}}$ is an $S^5$ overlap integral and
$\vev{V_{k_1}V_{k_2}V_{k_3}}$ is the 3-point string amplitude for the string states dual to the operators, with the {momenta}  $k_i$ depending on their dimensions and $R$-charges.  
The closed string coupling and tension are translated through the AdS/CFT dictionary to be  \mbox{$\gc=\pi^{3/2}/N$} and $\ap=1/\sqrt{\lambda}$, where $\lambda=g_{YM}^2N$ is the 't Hooft coupling.

In supergravity amplitudes the three-point couplings may have derivative terms, but this does not affect our results since our string amplitudes are on-shell and do not distinguish between derivative and non-derivative couplings.  As an illustration of this point,  the  difference between Witten diagrams for the supergravity couplings $\phi_1\phi_2\phi_3$ and $\phi_1\partial_\mu\phi_2\partial^\mu\phi_3$ is the factor \cite{Freedman:1998tz}
\be\label{Wittdiff}
\Delta_2\Delta_3+(d-2\Sigma)\alpha_1\approx \half(\Delta_1^2-\Delta_2^2-\Delta_3^2)\,,
\ee
where the approximation assumes that $\Delta_i\gg1$.   But this last factor is precisely $-k_2\cdot k_3$ where the $k_i$ correspond to the momentum along the $AdS_5$ directions in the flat-space limit \cite{Minahan:2012fh}.  Hence one should use the Witten diagram for non-derivative couplings since the derivative terms, if any,  will be built into $\GG_{123}$ in our analysis.  Further evidence that this is correct, at least to leading order, is  our ability to reproduce the supergravity results \cite{Lee:1998bxa} for three-point functions of chiral primaries in \cite{Bargheer:2013faa}, and our results we present in \secref{sec:spin} which among other things reproduce the supergravity results in \cite{Arutyunov:1999en}.

A prominent feature of (\ref{3corr1}) is  the presence of poles for  extremal correlators, which occur when one of the $\al_i$ is zero.  This corresponds to the dimension of one operator being equal to the sum of the other two.  In the case of three chiral primaries it was shown that the pole is canceled  by a zero in the coupling $\GG_{123}$, leaving a finite result \cite{Lee:1998bxa,D'Hoker:1999ea}.  However, if  the correlator contains one or more non-primary operators then this might not be the  case.  In particular, in \cite{Bargheer:2013faa}, the coupling for two chiral primaries with opposite $R$-charges and the Konishi operator was explicitly computed, where it was found that
\be\label{CCNC}
\vev{V_{C,J} V_{C,-J} V_K}\approx\frac{\gc^3}{16}\ap^2(J+\half\Delta)^4\,.
\ee
$V_{C,J}$ is the vertex operator for a chiral primary $\OO_J(x)$ with $R$-charge $J$ and $V_K$ is the vertex operator for the Konishi operator, with dimension $\Delta\approx 2\,\lambda^{1/4}$.
 As extremality is approached with $J\to{}_+\Delta/2$ it is clear that the coupling in (\ref{CCNC}) remains non-zero and the pole in (\ref{3corr1}) survives.
This indicates that the Konishi operator mixes with the double-traced $SO(6)$ singlet  in the tensor product of two $J$-symmetric traceless representations, with 
\be\label{dt}
 \OO_{J\bar J}(x)=\ :\OO_J(x)\OO_{- J}(x):
 \ee
being  one of the components that contribute equally to the mixing.

   In this paper we  compute the mixing for large $J$, and hence the splitting of the dimensions at the extremal point.  If two equal dimension operators mix, there will be a log divergence in the two-point function between the operators.  Our result  at closest approach of the splitting between the two eigenvalues of the mixed operators is 
 \be\label{splitting}
 \Delta m\approx \frac{2\sqrt{M}\,J^{3/2}}{N}\,,
 \ee
 where $M$ is the dimension of the $J$-symmetric, traceless representation of $SO(6)$,

In \secref{sec:review} we review results from \cite{Minahan:2012fh,Bargheer:2013faa} which are necessary for the analysis in this paper.  In \secref{sec:extreme} we present the details about the three-point correlators between the Konishi operator and two chiral primaries of opposite $R$-charge.  We then show how this leads to the mixing between Konishi and the double-trace operators.  In \secref{sec:spin} we consider the three-point correlators between scalar operators dual to supergravity states and states at the first massive level, and at least one higher spin operator dual to a string state along the leading Regge trajectory.   We show here that we can match with the results in \cite{Costa:2012cb} as long as the scalars also have large $R$-charges such that our flat-space approximations are valid.  In \secref{sec:conclusions} we present our conclusions.  We also include an appendix with further details.


\section{Review of previous results}
\label{sec:review}

In this section we collect some relevant results from~\cite{Minahan:2012fh,Bargheer:2013faa}.

We assume that the operators are short operators such that the sizes of the string-duals are small compared to the AdS$_5$ and S$^5$ radii. We can then approximate the three-point correlator  using Witten diagrams~\cite{Witten:1998qj,Freedman:1998tz}.
The Witten diagram includes an integration over all possible intersection
points, but in the case of large dimension operators the integration is dominated by a saddle point.  The location of the saddle point can be determined~\cite{Klose:2011rm,Minahan:2012fh} as well as the 
contribution of Gaussian fluctuations.  The saddle-point location is itself related to the conservation of constants of the motion, and it is these constants that determine the vertex operators that are inserted in the Witten diagram \cite{Minahan:2012fh}. 

The classical constants of the motion  are expectation values of the different components of the conformal algebra and are 
determined by the $R$-charges, the spins, the dimensions and by the
positions of the operators on the boundary.  For example,  for the scalar operator $\OO_i(x_i)$ in a three-point function the conserved charges include
\be\label{charges}
\vev{P^\mu_{i}}\equiv\vev{ \OO_j(x_j) \OO_k(x_k)P^\mu \OO_i(x_i)} \eq-2i\left(\alpha_k\frac{x_{ij}^\mu}{x_{ij}^2}+\alpha_j\frac{x_{ik}^\mu}{x_{ik}^2}\right)\,,\nn\\
\vev{D_i}\equiv\vev{ \OO_j(x_j) \OO_k(x_k)D \OO_i(x_i)} 
\eq i\biggbrk{\Delta_i-2\,x_{i\mu}\left(\alpha_k\frac{x_{ij}^\mu}{x_{ij}^2}+\alpha_j\frac{x_{ik}^\mu}{x_{ik}^2}}\right)
\,,
\ee
where $x^\mu_{ij}=x^\mu_i-x^\mu_j$.

The conformal algebra can be written in a manifestly
$\grp{SO}(2,d)$
covariant way by defining
\be\label{so2d}
M_{-1\mu}\equiv\sfrac{1}{\sqrt{2}}(\kappa P_\mu-\kappa^{-1}K_\mu)\,,\qquad M_{d\mu}\equiv\sfrac{1}{\sqrt{2}}(\kappa P_\mu+\kappa^{-1}K_\mu)\,,\qquad M_{-1d}\equiv -D\,,
\ee
with arbitrary $\kappa$, and Casimir $-\half M_{rs}M^{rs}=-\Delta^2$, $r,s=-1\dots d$. Setting
$d=4$ and combining with the   $\grp{SO}(6)$ $R$-symmetry Casimir, we have
\be
-\half M_{rs}M^{rs}+\half R_{IJ}R^{IJ}=-\Delta^2+J^2\,,
\ee
where $R_{IJ}$ are the $R$-symmetry generators with
$I,J=5,\dots,10$.

If $\Delta$ and $J$ are large, then the $\grp{SO}(2,4)\times
\grp{SO}(6)$ algebra effectively reduces to a 10 dimensional Poincar\'e algebra.
Because of the translation symmetry, we are free to shift the intersection point to $x^\mu=0$, in which case all $\vev{M_{\mu\nu}}=0$ and the conserved charges $\vev{K^\mu_i}$ satisfy
\be\label{KPrel}
\vev{K_i^\mu}=-\frac{\al_1\al_2\al_3\Sigma}{2F^2}x_{12}^2x_{23}^2x_{31}^2\vev{P_i^\mu}\,,
\ee
where
\be
F=\al_1\al_2\, x_{12}^2+\al_2\al_3\,x_{23}^2+\al_3\al_1\,x_{13}^2\,.
\ee
If we further choose $\kappa$ to be
\be\label{kappa}
\kappa =\frac{\sqrt{\al_1\al_2\al_3\Sigma}}{\sqrt2F}|x_{12}||x_{23}||x_{31}|\,,
\ee
then the only non-zero components in \eqref{so2d} are $\vev{M_{-1m}}$,
$m=0,\dots,4$
for all three operators.  Assuming $\Delta^2\gg1$, and choosing a
basis where the only non-zero $R$-symmetry components are $\vev{R_{J,10}}$, we can then identify the full 10-dimensional flat-space momentum as
\be\label{10dmom}
k^M=\bigbrk{\vev{M_{-1m}},\vev{R_{J,10}}}\,,
\ee
which satisfies the on-shell condition
\be\label{k2}
k\cdot k=-\Delta^2+J^2=-4n\sqrt{\la}\,.
\ee

We can expand the algebra to the full superconformal $\grp{PSU}(2,2|4)$,
by including the 
supercharges  $Q_{\al a}$ and $\tQ^a_{\dda}$ and the
superconformal generators $S^a_{\al}$ and $\tS_{\dda a}$, where $\al$ and
$\dda$ are space-time
spinor indices and  raised or lowered $a$ is an $\grp{SO}(6)$ spinor index.  The $\grp{SU}(2,2)\simeq\grp{SO}(2,4)$ covariant supergenerators are
\be
Q^1_{\dot a a}\eqv(\kappa^{1/2}Q_{\al a},\kappa^{-1/2}\tilde S_{\dot\al a})\nn\\
Q^{2,\dot a a}\eqv(\kappa^{-1/2}\eps^{\al\beta}S^a_{\beta},\kappa^{1/2}\eps^{\dot\al\dot\beta}\tilde Q^a_{\dot\beta})
\ee
where  lowered or raised $\dot a$ are $\grp{SO}(2,4)$ spinor indices.  Defining the supercharges
\be\label{QLQRdef}
Q^{\mathrm{L}}_A=Q^1_{\dot a a}+\gamma^{-1}_{\dot b\dot a}\gamma^{\ 6}_{ba}Q^{2,\dot bb}\,,\qquad
Q^{\mathrm{R}}_A=-i\bigbrk{Q^1_{\dot a a}-\gamma^{-1}_{\dot b\dot a}\gamma^{\ 6}_{ba}Q^{2,\dot bb}}\,,
\ee
then in the flat-space limit these approach the usual 10-dimensional
super-Poincar\'e generators with
\begin{equation}
\bigacomm{Q\supup{L,R}_A}{Q\supup{L,R}_B}
=
-2(\mathrm{P_+}\Gamma^M C)_{AB}P_M\,,
\qquad
P_M=(M_{-1m},R_{J,10})\,,
\qquad
\bigacomm{Q\supup{L}_A}{Q\supup{R}_B}
=
0\,,
\label{eq:poincare}
\end{equation}
 where $\mathrm{P_+}$ is the
positive-chirality projector.

A special class of operators are primary, which satisfy
$[S^a_\al,\OO(0)]=[\tS_{\dda a},\OO(0)]=0$.
In terms of the super-Poincar\'e generators this corresponds to
\be\label{twistcond}
Q\supup{L}_{\al \tilde a}=i\,Q\supup{R}_{\al \tilde a}\,,
\qquad
{Q\supup{L}_{\dot\al}}^{\tilde a}=-i\,{Q\supup{R}_{\dot\al}}^{\tilde a}\,,
\ee
where 
($\al,\dot\al$) are the explicit four-dimensional space-time spinor indices transverse to the trajectory  in the AdS$_5$ part and  ($\tilde a$) are the $SO(6)$ spinor
indices for the remaining six dimensions. 

The operators that satisfy (\ref{twistcond}) are described in detail in \cite{Bargheer:2013faa}.  They are linear combinations of operators in the NS-NS and R-R sectors and have a fairly complicated form.  However, their flat-space three-point functions were remarkably simple.  In particular, the three-point function for two chiral primaries and a Konishi or Konishi-like operator (a primary operator at level one with a nonzero $R$-charge), is given by
\be
\label{eq:wwv}
\vev{V_{C,J_1} V_{C,J_2} V_{K,J_3}}=\gc^3\,\ap^2\,\frac{\al_1^2\al_2^2\Sigma^4\tilde\al_3^2\tilde\Sigma^2}{\Delta_1^2\Delta_2^2\Delta_3^4}\,,
\ee
where 
\be
\tilde\Sigma=\sfrac{1}{2}(|J_1|+|J_2|+|J_3|)\,,\qquad\tilde\alpha_i=\tilde\Sigma-|J_i|\,.
\ee
In the special case where $J_1=-J_2=J$, $J_3=0$, this reduces to (\ref{CCNC}),
where $\Delta=\Delta_3$.


\section{Extremal correlators and operator mixing}
\label{sec:extreme}

In this section we compute the mixing between the double trace operator $\OO_{J\bar J}$ and the Konishi operator $\OO_K$.   We first observe that the coupling in (\ref{coupling}) near the extremal point $J=\Delta/2\approx\lambda^{1/4}$ is well approximated by 
\be
\GG_{123}\approx \frac{2^3\pi\sqrt{\la}}{ N}\,,
\ee
where we have used the AdS/CFT dictionary for $g_c$ and $\al'$. Using this, we also have that near the extremal point the three-point coupling coefficient in (\ref{3corr2}) is approximately
\be\label{extremalcorr}
\CC_{123}\approx \frac{\la^{3/8}}{N}\frac{1}{2J-\Delta}\,.
\ee

The structure constants among physical primary operators must be finite, hence the pole in (\ref{extremalcorr}) signals that the operators  renormalize and mix with each other.  To find the mixing 
the double trace operator $\OO_{J\bar J}(x)$ defined in (\ref{dt}) needs to be regulated. Here we use point-splitting and define the regulated operator
\be\label{rdt}
 \OO^\eps_{J\bar J}(x)=\ :\OO_J(x+\eps)\OO_{\bar J}(x):=\OO_J(x+\eps)\OO_{\bar J}(x)-\frac{1}{|\eps|^{2J}}\,.
 \ee
It is then clear that
\be
\langle \OO^\eps_{J\bar J}(x) \OO^\eps_{J\bar J}(y)\rangle=\frac{1}{|x-y|^{4J}}\,,
\ee
while the mixing in the two-point function is
\be\label{mixingJJ}
\langle \OO^\eps_{J\bar J}(x)\OO_K(y)\rangle&=&\langle\OO_J(x+\eps)\OO_{\bar J}(x)\OO_K(y)\rangle\nn\\
&=&\frac{\CC_{123}}{|x-y|^{2\Delta}|\eps|^{2J-\Delta}}\approx \frac{\la^{3/8}}{N}\frac{1}{(2J-\Delta)|\eps|^{2J-\Delta}}\frac{1}{|x-y|^{2\Delta}}\,.
\ee
Note that the planar three-point function gives a nonplanar contribution to the two-point function of the Konishi operator with the double trace operator. As expected, in the limit of infinite $N$ there is no mixing between these operators.

$\OO^\eps_{J\bar J}(x)$ is just one of $M$ operators that have an equal two-point function with $\OO_K(x)$, where $M$ is the dimension of the $J$-symmetric traceless representation of $SO(6)$,
\be
\label{dim}
M=\frac{(J+1)(J+2)^2(J+3)}{12}\approx \frac{J^4}{12}\,.
\ee
Since $\OO_K(x)$ is an $SO(6)$ singlet, it only mixes with the singlet combination of the double-trace operators,
\be\OO^\eps_{s,J}(x)=\frac{1}{\sqrt{M}}\sum_{I=1}^M\left(\OO_I(x+\eps)\OO_I(x)-\frac{1}{|\eps|^{2J}}\right)\,,
\ee
where $I$ refers to one of the states that make up the $J$-symmetric traceless representation.  The two-point-function of the singlet double-trace and the Konishi operators is then
\be\label{mixing}
\langle \OO^\eps_{s,J}(x)\OO_K(y)\rangle\approx \frac{\la^{3/8}\sqrt{M}}{N}\frac{1}{(2J-\Delta)|\eps|^{2J-\Delta}}\frac{1}{|x-y|^{2\Delta}}\,.
\ee
which shows, besides the pole, a logarithmic divergence in the UV cutoff. To cancel these, we take as the renormalized operators the combinations
\begin{align}
\OO^{\eps,ren}_{s,J} &= \OO^\eps_{s,J} +\lim_{\Delta\to2J}\frac{\lambda^{3/8}\sqrt{M}\log(\eps\mu)}{N}\mu^{2J-\Delta}\, \OO_{K}\,,\nn \\
\OO_K^{ren} &= \OO_K - \lim_{\Delta\to2J}\frac{\lambda^{3/8}\sqrt{M}}{N(2J-\Delta)}\mu^{\Delta-2J}\, \OO^\eps_{s,J}\,, 
\end{align}
where $\mu$ is some scale and the powers of $\mu$ are necessary to match dimensions.
Notice that to order $1/N$ the pole in the three-point function of $\OO_K^{ren}$  with the two chiral primaries $\OO_J$ and $\OO_{\bar J}$ cancels, leaving us with
\be\label{3pointren}
\langle\OO_J(x_1)\OO_{\bar J}(x_2)\OO_K^{ren}(x_3)\rangle\approx\frac{\la^{3/8}\sqrt{M}}{{N}|x_{13}|^{2J}|x_{13}|^{2J}}\log\frac{|x_{12}|}{\mu|x_{13}||x_{23}|}\,,
\ee
which does not have the usual form since $\OO_K^{ren}$ is not a primary operator at this order.
The pole in the three-point function of $\OO^{\eps,ren}_{s,J}$ with the same chiral primaries does not appear until order $1/N^2$, and so to leading order in the $1/N$ expansion it  is  also pole-free.

The anomalous dimension matrix at this order has the form 
\be
\Gamma=\left(\begin{array}{cc}0&\delta_{sJ,K}\\
\,\delta_{K,sJ}&0\end{array}\right)\,,
\ee
where the off-diagonal elements are given by \footnote{In a previous version of this paper equations (\ref{deltaeq})-(\ref{splitting2}) were off by a factor of $1/2$ which led to a mismatch with the recent result in \cite{Korchemsky:2015cyx}.}
\begin{align}\label{deltaeq}
\delta_{sJ,K}=&\mu\frac{\partial}{\partial\mu}\frac{\la^{3/8}\sqrt{M}}{N}\log(\eps\mu)= \frac{\la^{3/8}\sqrt{M}}{N}\,,\nn \\
\delta_{K,sJ}=&\mu\frac{\partial}{\partial\mu}\left( - \lim_{\Delta\to2J}\frac{\lambda^{3/8}\sqrt{M}}{N(2J-\Delta)}\mu^{\Delta-2J}\right)=\frac{\la^{3/8}\sqrt{M}}{N}\,.
\end{align}
The eigenvalues of $\Gamma$ are then given by
\be
\delta_{\pm}=\frac{\lambda^{3/8}}{N}\,\sqrt{M}\,.
\ee
Hence, at the cross-over point $J=\Delta/2\approx \lambda^{1/4}$, the mixed operators have dimensions
\be\label{newdim}
\Delta_{\pm}=2J\pm\frac{\sqrt{M}\lambda^{3/8}}{N}
\ee
and thus the splitting is $\Delta_+-\Delta_-$, giving the result in \eqref{splitting}, which can be reexpressed as
\be\label{splitting2}
 \Delta m\approx \frac{2\sqrt{M}\,\lambda^{3/8}}{N}\approx\frac{\lambda^{7/8}}{\sqrt{3}\,N}\,.
 \ee
Note that the structure constant in (\ref{3pointren}) is scheme dependent as we could have defined different renormalized operators by using other linear combinations of the bare operators.  But this will not affect the structure constants between the chiral primaries and the final mixed operators that are the eigenstates of $\Gamma$.
 
We can compare the splitting to the leading   correction to the double-trace dimension  coming from supergravity, where it was found to be
\cite{Dolan:2001tt}
\be\label{djj}
\delta_{JJ}=-\frac{2 (J-1)J(J+2)}{N^2}\,.
\ee
Hence the splitting, which scales as $J^{7/2}/N$, is much larger.



\section{Three-point couplings with spin}
\label{sec:spin}


\subsection{The operators and corresponding string states}
In this section we  consider three-point functions where at least one of the operators is dual to a string state in the leading Regge trajectory. As in \cite{Minahan:2012fh,Bargheer:2013faa}, the correlators can have one or more chiral primaries,
\be
\OO\indup{CP}=C_{I_1I_2\dots I_J}\tr[\Phi^{I_1}\Phi^{I_2}\dots\Phi^{I_J}]\,,
\ee
where $C_{I_1I_2\dots I_J}$ is symmetric and traceless and $\Phi^I$ are the six scalars in \sym.  These are states in the $[0,J,0]$ representation of $\grp{SO}(6)$ and have the protected dimension $\Delta\indup{CP}=J$. At strong coupling, these operators correspond to massless scalar vertex operators with nonzero Kaluza-Klein momentum, and have the following mixture of NS-NS and R-R modes \cite{Minahan:2012fh,Bargheer:2013faa}
\be
\label{VChiral}
V_{C,k}=-\sfrac{1}{4}\bigbrk{W_{1,k}+\sfrac{1}{\sqrt{2}} W_{2,k}}\,.
\ee
The NS-NS and R-R vertices are given respectively by
\be
\label{VMassless}
W_{1,k}=\gc\,\ve_{M\tilde M}\,\psi^Me^{-\phi}\tilde\psi^{\tilde M}e^{-\tilde\phi}e^{ik\cdot X}\,,
\qquad
W_{2,k}=\gc\,t_{AB}\,\tilde\Theta^Ae^{-\half\tilde\phi}\,\Theta^Be^{-\half\phi}e^{ik\cdot X}\,,
\ee
with the following polarizations
\be
\label{PolarMassless}
\ve_{M\tilde M}\eq(-1)^{\s_k(\tilde M)}\left(\eta_{M \tilde M} - \frac{k_M \bar k_{\tilde M}+k_{\tilde M} \bar k_M}{k\cdot \bar k}\right),\; t_{AB}=\left(\sapt\right)^{1/2}
\bigbrk{C^{\dagger}i\Gamma^{0'}\Gamma^{1'}\Gamma^{2'}\Gamma^{3'}\slk}_{AB}\,.
\ee
We define the untwisting factor as
\be
\sigma_k(M)=
\begin{cases}
1 & M=0',\dots,3'\\
0 & M=4',5,\dots,9\,.
\end{cases}
\ee
The primed indices denote a frame in which the directions $0',\ldots,3'$ are transverse to the momentum $k$.

The other operators considered in \cite{Minahan:2012fh,Bargheer:2013faa} also transform in the $[0,J,0]$ representation, but  have bare dimension $\Delta_0=J+2$~\cite{Beisert:2002tn}. Operators of this type include 
\be
\OO_J=\tr[\Phi^I\Phi_I Z^J]+\dots\,,
\ee
 where $Z=\brk{\Phi_5+i\Phi_6}/\sqrt{2}$, and the ellipsis refers to different positions of the $\Phi_I$ in the trace, such that the corresponding magnon momenta lie at level one~\cite{Beisert:2002tn,Minahan:2002ve}. For $J=0$ we have the Konishi operator %
\be
\OO\indup{K}=\OO_{J=0}=\tr[\Phi^I\Phi_I]\,,
\ee
which at strong coupling has dimension $\Delta\indup{K}\approx2 \lambda^{1/4}$.
These operators correspond at strong coupling to scalar vertex operators at the first massive string level, possibly with some Kaluza-Klein momentum. They can be written as
\be
\label{VKonishi}
V_{K,k}=-\sfrac{1}{16}\bigbrk{V_{1,k}+V_{2,k}+\sfrac{1}{\sqrt 2} V_{3,k}}\,,
\ee
where $V_{1,k}$ and $V_{2,k}$ are the NS-NS vertices and $V_{3,k}$ is in the R-R sector
\begin{align}
\label{VMassive}
V_{1,k}&=\gc\stap\eps_{MN;\tM\tN}\,\psi^M i\p X^N e^{-\phi}\,\tilde\psi^{\tM}i\bar\p X^{\tN} e^{-\tilde\phi}e^{ik\cdot X}\,,
\nn\\
V_{2,k}&=\gc\,\al_{MNL;\tM\tN\tL}\,\psi^M\psi^N\psi^Le^{-\phi}\,\tilde\psi^\tM\tilde\psi^\tN\tilde\psi^\tL e^{-\tilde\phi}e^{ik\cdot X}\,,
\nn\\
V_{3,k}&=\frac{2\gc}{\alphap}
\big(i\bar\p X^M\tilde\Theta-\sfrac{\alphap}{16}\tilde\psi^M(\slk\tilde{\slpsi}\tilde\Theta)\big)^Ae^{-\tilde\phi/2}t_{MA;NB}\big(i\p X^N\Theta-\sfrac{\alphap}{16}\psi^N(\slk\slpsi\Theta)\big)^Be^{-\phi/2}
e^{ikX}\,.
\end{align}
Defining the tensor $\hat\eta^{MN}\equiv\eta^{MN}-\frac{k^M k^N}{k^2}$, we can write the polarizations in the following way
\begin{align}
\label{PolarMassive}
\eps_{MN;\tM\tN}&= (-1)^{\s_k(\tilde M)+\s_k(\tilde N)}\left(
\sfrac12(\hat\eta_{M\tM}\hat\eta_{N\tN}+\hat\eta_{M\tN}\hat\eta_{N\tM})-\sfrac{1}{9}\hat\eta_{MN}\hat\eta_{\tM\tN}\right)\,,\nn\\
\al_{MNL;\tM\tN\tL}&=(-1)^{\s_k(\tilde M)+\s_k(\tilde N)+\s_k(\tilde L)}
\sfrac{1}{3!^2}\bigbrk{\hat\eta_{M\tM}\hat\eta_{N\tN}\hat\eta_{L\tL}\mp(5\text{ permutations})}\,,\nn\\
t_{MA;NB}&=(-1)^{\s_k(\tilde M)}\left(\sapt\right)^{1/2}\bigbrk{C^\dagger i\Gamma^{0'}\Gamma^{1'}\Gamma^{2'}\Gamma^{3'}\slk(\hat\eta_{MN}- \sfrac19\Gamma^R\Gamma^S\hat\eta_{MR}\hat\eta_{NS})}_{AB}\,.
\end{align}
In this work we will also consider relatives of the Lagrangian operator with $R$-charge $J$ which have protected dimension $\Delta_{\op L_J}=4+J$
\be
\OO_{\op L_J}=\tr[F_{\mu\nu} F^{\mu\nu} Z^J]+\dots \,,
\ee
where $F_{\mu\nu}$ is the field strength of \sym, and the ellipsis refers to different positions of the $F_{\mu\nu}$ in the trace, and also other scalar and spinor terms required by supersymmetry. At strong coupling these operators correspond to the ten-dimensional dilaton with some Kaluza-Klein momentum
\be
\label{VDilaton}
V_{\op L,k}= \frac{\gc}{\sqrt 8} \left(\eta_{M \tilde M} - \frac{k_M \bar k_{\tilde M}+k_{\tilde M} \bar k_M}{k\cdot \bar k}\right) \psi^{M} e^{-\phi} \tilde\psi^{\tilde M} e^{-\tilde\phi} e^{i k\cdot X}\,.
\ee

Finally, an obvious extension to the results in \cite{Bargheer:2013faa} is to consider operators with spin. It is in general nontrivial to find the vertex operator for a generic operator with dimension $\Delta$ and spin $S$. However, if the operator is a special combination of the three following twist two operators \cite{Costa:2012cb}
\be
\tr[F_{\mu\nu_1}D_{\nu_2}\ldots D_{\nu_{S-1}}{F_{\nu_S}}^\mu]\,, \quad \tr[\Phi_{I}D_{\nu_1}\ldots D_{\nu_S}\Phi^I]\,,\quad\tr[\bar\psi_\alpha D_{\nu_1}\ldots D_{\nu_{S-1}}\Gamma_{\nu_S}\psi^\alpha]\,,
\ee
then the dual string  state lies on the leading Regge trajectory at level $n=\sfrac{1}{2}(S-2)$  with scaling dimension $\Delta\approx2\sqrt n \lambda^{1/4}$. 

As noted in \cite{Schlotterer:2010kk}, the vertex operator for a symmetric traceless state in the leading Regge trajectory is given by
\be
O_{n,k}=\gc\left(\stap\right)^n\eps_{M_1 \ldots M_{n+1} \tilde M_1 \ldots \tilde M_{n+1}} \prod_{j=1}^n \bigl(i\partial X^{M_j}\bigr) \psi^{M_{n+1}} e^{-\phi}
\prod_{j=1}^n \bigl(i\bar \partial X^{\tilde M_j}\bigr) \tilde\psi^{\tilde M_{n+1}} e^{-\tilde\phi} e^{i k\cdot X} \,,
\ee
with $\eps_{M_1 \ldots M_S}$ a totally symmetric and traceless tensor.
To normalize the vertex operator we need only to compute its two-point function
\be
\vev{O_{n,k} O_{n,-k}}= \Gamma(S/2)^2 \eps_{M_1\ldots M_S} \eps^{M_1\ldots M_S}\,.
\ee
The normalized vertex operator for a string state in the leading Regge trajectory is then
\be
\label{VRegge}
V_{S,k}= \frac{\gc 2^{S/2-1}}{\Gamma(\sfrac{S}{2}){\ap}^{S/2-1}} \eps^{norm}_{M_1 \ldots M_{S/2} \tilde M_1 \ldots \tilde M_{S/2}} \prod_{j=1}^{\frac{S-2}{2}} \bigl(i\partial X^{M_j}\bigr) \psi^{M_{S/2}} e^{-\phi}
\prod_{j=1}^{\frac{S-2}{2}}  \bigl(i\bar \partial X^{\tilde M_j} \bigr) \tilde\psi^{\tilde M_{S/2}} e^{-\tilde\phi} e^{i k\cdot X} \,.
\ee


\subsection{The building blocks}
Three-point functions of operators with arbitrary spin can be written as a linear combination of a finite number of conformally invariant building blocks \cite{mack1977,Sotkov1977375,Osborn:1993cr}, which have been rederived in \cite{Costa:2011mg} using the embedding formalism
\be
\label{spincorr}
 C(\{x_i\}) = \frac{Q(\{x_i\})}{(x_{12}^2)^{\sfrac{\tau_1+\tau_2-\tau_3}{2}}(x_{13}^2)^{\sfrac{\tau_1+\tau_3-\tau_2}{2}}(x_{23}^2)^{\sfrac{\tau_2+\tau_3-\tau_1}{2}}} \,,
\ee
where $\tau_i= \Delta_i + S_i$ and $Q(\{x_i\})$ is a polynomial built out of six building blocks
\be
\label{blocks}
V_1 \equiv V_{1,23} \,, \qquad V_2 \equiv V_{2,31} \,, \qquad V_3 \equiv V_{3,12} \,, \qquad H_{12} \,, \qquad H_{13} \,, \qquad H_{23} \,,
\ee
which encode the space-time dependence 
\be
V_{i,jk}  =  \frac{1}{x_{jk}^2} \left(x_{ij}^2  (z_i\cdot x_{ik}) -x_{ik}^2  (z_i\cdot x_{ij})  \right) \,, \qquad H_{ij}  =  x_{ij}^2 (z_i\cdot z_j) -2 (z_i\cdot x_{ij})(z_j\cdot x_{ij}) \,.
\ee
Here we encode symmetric tensors by polynomials obtained by contracting the tensor with a reference vector $z_i^\mu$. We can recover the tensor by applying the symmetric traceless projector
\be
P_{\mu_1 \ldots \mu_S}^i = \frac{1}{(S!)^2} D^i_{\mu_1} \ldots D^i_{\mu_S}\,, \qquad D^i_\mu=\left(1+z_i\cdot \frac{\partial}{\partial z_i}\right)\frac{\partial}{\partial z_i^\mu}-\frac{1}{2}{z_i}_\mu \frac{\partial^2}{\partial z_i\cdot \partial z_i}\,.
\ee
The polynomial $Q(\{x_i\})$ is then written as the following sum
\be
\label{polynomial}
P_{\mu_1 \ldots \mu_{S_1}}^1 P_{\nu_1 \ldots \nu_{S_2}}^2 P_{\rho_1 \ldots \rho_{S_3}}^3 \sum_{i,j,k \in \mathcal I} C_{ijk}V_1^{S_1-i-j} V_2^{S_2-i-k} V_3^{S_3-j-k} H_{12}^{i} H_{13}^{j} H_{23}^{k} \,,
\ee
where $C_{ijk}$ are the structure constants and the summation range is
\be
\label{summationrange}
\mathcal I = \left\{ i,j,k \in \mathbb N_0 : S_1-i-j \geq 0, \quad S_2-i-k \geq 0, \quad S_3-j-k \geq 0\right\} \,.
\ee

\begin{figure}[h]
\centering
\def\svgwidth{0.8\columnwidth} 
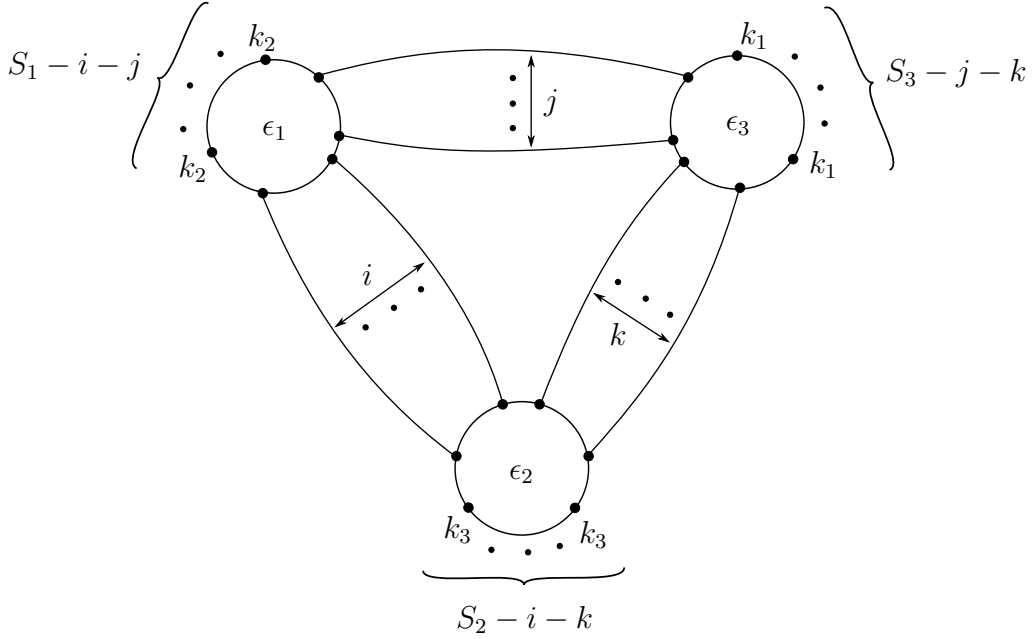
\caption{We represent each polarization $\epsilon_i$ by a circle with $S_i$ indices. The lines connecting two polarization tensors correspond to contractions of their indices. The leftover indices are contracted with the momenta. The allowed structures correspond to the different ways of connecting the three polarization tensors.}
\label{fig:drawing}
\end{figure}

We can reproduce these structures with our holographic procedure. In \cite{Schlotterer:2010kk} Schlotterer derived the coupling of three Regge open superstring states at levels $n_1$, $n_2$ and $n_3$.\footnote{This work is an extension of \cite{Sagnotti:2010at} which addresses a related problem in the context of open bosonic strings.} We can apply this result directly to the left and right movers separately. The three-point function for the left movers is,
\begin{align}
\label{3Regge}
\langle O_{n_1} O_{n_2} O_{n_3}\rangle_L =& \sqrt{\alphap/2}^{s_1+s_2+s_3}n_1! n_2! n_3! \sum_{i,j,k\in \mathcal I}\frac{ (\alphap/2)^{-i-j-k}(i s_3 +j s_2 +k s_1 -ij-ik-jk)}{i!j!k!(s_1-i-j)!(s_2-i-k)!(s_3-j-k)!} \nn\\
&\times (\epsilon^1\cdot k_2^{s_1-i-j}) (\epsilon^2\cdot k_3^{s_2-i-k}) (\epsilon^3\cdot k_1^{s_3-j-k})\eta_{12}^i \eta_{13}^j \eta_{23}^k \,,
\end{align}
where $s_i=n_i+1$.  We used the following shorthand notations
\be\epsilon^i \cdot k_j^p &=& \epsilon^i_{M_1 \ldots M_p  M_{p+1}\ldots M_{s_i}}k_j^{M_1}\ldots k_j^{M_p}\,,\nn\\
\epsilon^i \epsilon^j \eta^k_{ij}&= &\epsilon^i_{M_1 \ldots M_{s_i}} \epsilon^j_{N_1 \ldots N_{s_j}} \eta^{M_1 N_1}\ldots\eta^{M_k N_k}\,,
\ee
and the summation range is the same as in \eqref{summationrange}.
We notice then that the superstring amplitude of three closed spin states with polarizations $\varepsilon_1, \varepsilon_2 \mbox{ and }\varepsilon_3$ will be a sum of terms of the form
\be
\label{blocksS}
(\epsilon^1\cdot k_2^{S_1-i-j}) (\epsilon^2\cdot k_3^{S_2-i-k}) (\epsilon^3\cdot k_1^{S_3-j-k})\eta_{12}^i \eta_{13}^j \eta_{23}^k \,,
\ee
which we represent diagrammatically in \figref{fig:drawing}.
These superstring structures from the coupling of the three string states at the intersection point in the bulk are in fact related to the boundary CFT building blocks.

To understand this relation between the two sets of building blocks, it will be enough to restrict ourselves to the $AdS_3$ space defined by the bulk direction and the plane on the boundary defined by the three endpoints of the strings. If the polarization is orthogonal to any of those directions then $\varepsilon^i \cdot k_j$ vanishes trivially and $\varepsilon^i \cdot \varepsilon^j$ remains proportional to $H_{ij}$.
On the other hand, if the polarizations live in some direction inside the $AdS_3$ defined above then we must be more careful. At the intersection point the polarization of each string is orthogonal to the momentum which, following \cite{Minahan:2012fh}, is given by
\be
k^\mu= \sfrac{1}{\sqrt 2} (\kappa \langle P^\mu \rangle - \kappa^{-1} \langle K^\mu\rangle) \,, \qquad k^z = - \langle D \rangle \,,
\ee
with  $\langle P^\mu\rangle$, $\langle D\rangle$, $\langle K^\mu\rangle$ and  $\kappa$ defined in \eqref{charges}, \eqref{KPrel}, and \eqref{kappa} when one takes the intersection point to the origin. In this case, the momentum at the intersection point can be written in the following way
\be
k_i^\mu = -2\frac{\sqrt{\al_1\al_2\al_3\Sigma}  |x_{12}||x_{23}||x_{13}|}{F}  \left(\alpha_k\frac{x_{ij}^\mu}{x_{ij}^2}+\alpha_j\frac{x_{ik}^\mu}{x_{ik}^2}\right)\,, \qquad k_i^z = \Delta_i - 2 \frac{\alpha_j \alpha_k\Sigma x_{jk}^2}{F}\,.
\ee

Since the polarization is orthogonal to the momentum at the intersection point, it is useful to change to a basis $\{\vec O_i, \vec P_i\}$ in the plane transverse to the momentum. If we let $\vec k^B_i$ be the projection of $\vec k_i$ to the boundary then we can write the basis vectors as
\be
\vec O_i &= \vec k_i\times \vec k^B_i \,,  \qquad\qquad \vec P_i &= \vec k_i \times \vec O_i \,.
\ee
We denote this change of basis by $M_i^I$. When one looks at the propagation of the string from the intersection point to the boundary, one sees that $\vec{O}_i$ stays the same, but $\vec{P}_i$ becomes proportional to  $\vec k^B_i$. Finally, we return from the basis $\{\vec O_i, \vec k^B_i\}$ to the canonical one at the boundary through another change of basis $M_i^B$. The relation between the polarization at the intersecton point and at the boundary is then given by
\be
(\varepsilon^I_i)^M = {(M_i^I M_i^B)^M}_\mu \,(\varepsilon_i^B)^\mu\,,
\ee
where we denoted the polarization at the intersection point by $\varepsilon^I_i$ and the one at the boundary by $\varepsilon^B_i$. The polarization $\varepsilon^I_i$ is symmetric and traceless, which leads to a polarization at the boundary with the same properties.  $\varepsilon^B_i$ will then act as a symmetric and traceless projector ensuring the right tensor structure of $Q(\{x_i\})$.
 The structure at the intersection point $\varepsilon^I_i \cdot k_j$ then corresponds to one of the CFT building blocks at the boundary 
\be
\varepsilon^I_i \cdot k_j = (\varepsilon^B_{i})^\mu \left(- 2 \frac{\sqrt{\alpha_1 \alpha_2 \alpha_3 \Sigma} x_{jk}^2}{|x_{12}||x_{13}||x_{23}|\Delta_i} P^i_\mu \,V_i\right)\,.
\ee
Analogously, the structure $\varepsilon^I_i \cdot \varepsilon^I_j$ becomes a combination of the CFT building blocks at the boundary
\be
\varepsilon^I_i  \varepsilon^I_j \eta_{ij}= (\varepsilon^B_i)^{\mu} \frac{1}{x_{ij}^2}\,P^i_\mu P^j_\nu \left(H_{ij}+2\frac{\alpha_i\alpha_j}{\Delta_i \Delta_j} V_i V_j \right)  (\varepsilon^B_j)^{\nu} \,.
\ee
Putting all the elements together, terms of the form \eqref{blocksS} now become proportional to
\be
\label{stpol}
\underset{k\neq i,j}{\prod_{i<j}} (x_{ij}^2)^{-\sfrac{S_i+S_j-S_k}{2}}\prod_i V_i^{S_i-\sum_{j \neq i} n_{ij}} \prod_{i<j} \left(H_{ij}+2\frac{\alpha_i\alpha_j}{\Delta_i \Delta_j} V_i V_j \right)^{n_{ij}} \,.
\ee
 Notice that the spacetime dependence from the string theory amplitude combines with the one obtained from the propagation of the strings in AdS, $(x_{12}^{2})^{-\alpha_3}(x_{13}^{2})^{-\alpha_2}(x_{23}^{2})^{-\alpha_1}$, to produce the spacetime behaviour expected from conformal symmetry in \eqref{spincorr}.

It will also be useful to work out the case where one  operator is a scalar, for which we set $S_1=0$. With no loss of generality we choose $S_2\leq S_3$, so the amplitude will reduce to a sum over $(S_2+1)$ structures with coefficients $\mathcal F_k$
\be
\label{twoS}
\sum_{k=0}^{S_2} (\ap/2)^{\frac{S_2}{2}+\frac{S_3}{2}-k} (\epsilon^2\cdot k_3^{S_2-k}) (\epsilon^3\cdot k_1^{S_3-k})\eta_{23}^k \mathcal F_k \,.
\ee 
Following our analysis on the equivalence of the two sets of building blocks, the three-point function of a scalar with two spin states becomes
\begin{equation}
\mathcal D\sum_{k=0}^{S_2} \frac{(2\ap\alpha_1\alpha_2\alpha_3\Sigma)^{\frac{S_2}{2}+\frac{S_3}{2}-k}}{\Delta_2^{S_2-k}\Delta_3^{S_3-k}}V_2^{S_2-k} V_3^{S_3-k} \bigl(H_{23}+\sfrac{2\alpha_2\alpha_3}{\Delta_2\Delta_3}V_2 V_3\bigr)^k \mathcal F_k =\sum_{k=0}^{S_2}  C_k V_2^{S_2-k} V_3^{S_3-k} H_{23}^k \,,
\end{equation}
where $\mathcal D$ denotes both the contribution from the propagation of the strings in AdS \eqref{3corr2} and the coupling \eqref{coupling}
\be
\mathcal D= \frac{\sqrt{(\Delta_1-1)(\Delta_2-1)(\Delta_3-1)}}{2^{5/2}\pi}\,\frac{\Gamma(\al_1)\Gamma(\al_2)\Gamma(\al_3)\Gamma(\Sigma-2)}{\Gamma(\Delta_1)\Gamma(\Delta_2)\Gamma(\Delta_3)}\,\frac{8\pi}{\gc^2\ap}\,\vev{\psi_{J_1}\psi_{J_2}\psi_{J_3}}\,,
\ee
and the structure constants are
\be
\label{structuretwoS}
C_k =\mathcal D  \frac{(2\ap\alpha_1\alpha_2\alpha_3\Sigma)^{S_2/2+S_3/2-k}}{\Delta_2^{S_2-k}\Delta_3^{S_3-k}} \sum_{l=k}^{S_2} \binom{l}{k} (\ap\alpha_1\Sigma)^{k-l} \mathcal F_l \,.
\ee

In the other case of interest, we  let two of the operators be scalars, hence $S_1=S_2=0$. There is now only the structure $(\ap/2)^{S_3/2}(\epsilon^3\cdot k_1^{S_3}) \mathcal F$, with the coefficient $\mathcal F$ 
coming from the superstring amplitude.  The structure constant is then given by
\be
\label{structureoneS}
C= \mathcal D  \frac{(2\ap\alpha_1\alpha_2\alpha_3\Sigma)^{S_3/2}}{(\Delta_3)^{S_3}} \mathcal F \,.
\ee


\subsection{Structure constants}
Now that we have the necessary ingredients, we can compute the three-point coupling for different string states.


\subsubsection*{Two massless scalars and one Regge spin}
We start by computing the three-point functions 
having two massless scalars, which we can compare with previous results in the literature. When the two massless scalars are dilatons, we can 
insert the polarizations defined in \eqref{VDilaton} into the results of \appref{sec:contractions}, thus giving the string amplitude between two dilatons and one Regge spin
\be
\label{DDSamp}
\vev{V_{\op L} V_{\op L} V_{S}}= \frac{(2\ap\alpha_1 \alpha_2 \alpha_3 \Sigma)^{S/2}}{(\Delta_3)^S\Gamma(S/2)} V_3^S\,. 
\ee

Analogously, if we consider the massless scalars to be two chiral primaries, then we use the polarizations defined in \eqref{PolarMassless}, obtaining the following amplitude for two chiral operators with $R$-charges $J_1$ and $J_2$ and one Regge spin with some $R$-charge $J_3$
\be
\label{CCSamp}
\vev{V_{C} V_{C} V_{S}}=\frac{\tilde\alpha_3^2\tilde\Sigma^2(2\ap\alpha_1 \alpha_2 \alpha_3 \Sigma)^{S/2}}{J_1^2 J_2^2(\Delta_3)^S\Gamma(S/2)} V_3^S\,.
\ee
When the two chirals have opposite R-charge and the Regge spin has no $R$-charge, the expression \eqref{CCSamp} becomes \eqref{DDSamp}. In this case we can put this result into \eqref{3corr2} and \eqref{coupling},  and taking the asymptotics for large 't Hooft coupling, we 
find the structure constant for two massless scalars with dimension $\Delta$ and a Regge state with spin $S$ {and dimension $\Delta_S\approx\sqrt{2(S-2)}\lambda^{1/4}$,} 
\be
C_{ \op L\op L S} =\frac{(-1)^{\Delta}\pi^{3/2}(S-2)^{(S-3)/2+\Delta}\lambda^{-1/4 + \Delta/2}}{ 2^{S+\Delta-7/2}\Gamma(S/2)\Gamma(\Delta-1)\Gamma(\Delta)N} 2^{-\Delta_S}  \csc\left(\pi\Delta_S/2\right) \,.
\ee
This reproduces the results obtained in \cite{Costa:2012cb}, where the authors used 
Mellin transforms. Note also the presence of poles in the factor $ \csc\left(\pi\Delta_S/2\right)$ which are related to mixing with double-trace operators as discussed in the previous section.

Another case we consider is when both scalars have the same dimension $\Delta$, but with the Regge spin also at the massless level, which corresponds to a relative of the graviton with some $R$-charge $J_g$. When $1\ll | J_g| \ll \Delta$ then \eqref{CCSamp} becomes simply \eqref{DDSamp} and the structure constant we obtain in that limit is
\be
C_{CCG}= \frac{\sqrt\pi}{2^{3+J}N} \Delta \,,
\ee
again obtaining agreement with \cite{Costa:2012cb} for large scaling dimensions.

 If we wish to compute the coupling between three states with $R$-charges of the same magnitude, then we can compare with \cite{Arutyunov:1999en} where the authors used supergravity to compute the coupling between two chirals and a graviton with some Kaluza-Klein momentum. For chirals with dimension $J$ and a graviton with $R$-charge $J_g=J$ we get
\be\label{chichigrav}
C_{CCG}= \frac{J^{3/2}}{8N} \langle C^{J}C^{J}C^{J}\rangle\,,
\ee
where $\langle C^{J_1} C^{J_2} C^{J_3}\rangle$ is the unique $SO(6)$ invariant that can be formed from the $C^{J_i}$ appearing in the definition of the spherical harmonics on $S^5$
\be
\psi_{J}=\frac{2^{\frac{J-1}{2}}\sqrt{(J+1)(J+2)}}{\pi^{3/2}} C^J_{i_1\ldots i_J} x^{i_1}\ldots x^{i_J}\,.
\ee
The structure constant in \eqref{chichigrav}  agrees with \cite{Arutyunov:1999en} in the limit of large scaling dimensions.
For completeness we have also computed the coupling between one dilaton operator, a chiral and an operator in the leading Regge trajectory and have obtained a vanishing structure constant at leading order.


\subsubsection*{One massive scalar, one massless scalar and one Regge spin}
If we consider the three-point function of a Konishi operator, a dilaton with some $R$-charge $J$ and a Regge spin with the opposite $R$-charge,  we can use the results of \appref{sec:contractions} and plug in the polarizations from \eqref{PolarMassive,VDilaton}, obtaining the amplitude
\be
\vev{V_K V_{\op L} V_{S}}=  -\frac{(S-2)}{2^5\sqrt 2 } \frac{(2\ap\alpha_1 \alpha_2 \alpha_3 \Sigma)^{S/2}}{(\Delta_3)^S\Gamma(S/2)} V_3^S \,.
\ee
Notice that this amplitude vanishes when we take the Regge spin to $S=2$.
Going to the limit of large 't Hooft coupling, the structure constant in the case of a Regge spin at the first mass level is
\be
C_{K\op L S}=-\frac{J^4 \sqrt\pi}{2^{9+J} \sqrt 2 N} \lambda^{-3/4} \,.
\ee

If instead of the dilaton we consider a chiral primary, then we use the polarizations of \eqref{PolarMassless,PolarMassive} in the results of \appref{sec:contractions}. When the spin operator has $S=2$, the structure constant also vanishes at leading order.
When considering a Konishi, one chiral with $R$-charge $J$ and a massive Regge spin the amplitude for $J \ll \lambda^{1/4}$ is
\be
\vev{V_{K} V_{C} V_{S}}=  \frac{(S-2)}{2^6 } \frac{(2\ap\alpha_1 \alpha_2 \alpha_3 \Sigma)^{S/2}}{(\Delta_3)^S\Gamma(S/2)} V_3^S \,.
\ee
{When the Regge spin is at the first massive level we obtain}
\be
C_{K\op C S}=\frac{J^4 \sqrt\pi}{2^{6+J}  N} \lambda^{-3/4} \,.
\ee


\subsubsection*{Two massive scalars and one Regge spin}
The two massive scalars we 
consider here are the Konishi-like operators 
with the vertex operators \eqref{VKonishi}. To compute the correlation function between two of this operators and one Regge spin we use the results derived in \appref{sec:contractions} and plug in the polarization \eqref{PolarMassive} thus obtaining, in the case of a massive spin, where all operators can be taken to have no $R$-charge
\be
\vev{V_{K} V_{K} V_{S}}= \frac{(S+14)^2}{2^8  } \frac{(2\ap\alpha_1 \alpha_2 \alpha_3 \Sigma)^{S/2}}{(\Delta_3)^S\Gamma(S/2)} V_3^S\,.
\ee
{In the limit of large 't Hooft coupling, we get for the structure constant of two Konishi operators with dimension $\Delta_K\approx2\lambda^{1/4}$ and a massive Regge spin of dimension $\Delta_S$ satisfying the triangle inequalities}
\be
C^t_{KKS} = \frac{(10-S)^{\frac{S-1}{2}}(S+14)^2 \sqrt\pi \lambda^{\frac{1}{4}}}{2^{\frac{7}{2}+S}N \Gamma(S/2)} 2^{-2\Delta_K-\Delta_S} (2-\sqrt n)^{\Delta_K-\Delta_S/2}  (2+\sqrt n)^{-2+\Delta_K+\Delta_S/2} \,,
\ee
where $n$ is the mass level of the Regge spin. If the massive Regge spin 
does not obey the triangle inequalities 
then the structure constant acquires  poles, 
\be
C^{nt}_{KKS} =C^t_{KKS} \times \frac{\csc((\sqrt n-2)\pi\lambda^{1/4})}{2} \,.
\ee
When the Regge spin is at the massless level
and has an $R$-charge $J\ll\lambda^{1/4}$,  we get
\be
\vev{V_{K} V_{K} V_{G}}= \frac{(2\ap\alpha_1 \alpha_2 \alpha_3 \Sigma)^{S/2}}{(\Delta_3)^S\Gamma(S/2)} V_3^S\,.
\ee
Then the structure constant of a Konishi operator, a Konishi-like operator with some $R$-charge $J$ and a relative of the graviton with the opposite $R$-charge is
\be
C_{KKG}=\frac{\sqrt\pi}{2^{2+J}N}\lambda^{1/4}\,,
\ee
in the limit of large dimensions.


\subsubsection*{One massless scalar and two Regge spins}
When there are two spins with $S_2 \leq S_3$, there are $(S_2+1)$ different structure constants in \eqref{twoS}. In \appref{sec:contractions} we have computed the amplitude for a dilaton with two spins \eqref{DSS}, which we can rewrite as
\be
\langle V_{\op L} V_{S_2} V_{S_3}\rangle  =\sum_{k=0}^{S_2} (\alphap/2)^{S_2/2+S_3/2-k}  (\epsilon^2\cdot k_3^{S_2-k})(\epsilon^3\cdot k_1^{S_3-k})\eta_{23}^{k} \, \mathcal D_k \,,
\ee
with the coefficients given by 
\be
\mathcal D_k =\frac{1}{\sqrt 8} \frac{1}{\prod_{i=2}^3(\frac{S_i-2}{2})!(\frac{S_i}{2})_{-\lfloor Y\rfloor-1}} \sum_{j=0}^{\lfloor Y\rfloor}\frac{2^{2-2Y}(2j+2)!}{j!(j+1)! (2Y-2j)!} \prod_{i=2}^3\frac{\left(\frac{S_i}{2}-j-2\right)_{j+1-\lfloor Y\rfloor}}{\left(S_i-2j-4\right)_{2j-2Y+1}}\,,
\ee
{where $Y=k/2-2$, $\lfloor Y\rfloor$ denotes the largest integer not greater than $Y$ and $(a)_m$ is the Pochhammer symbol, $(a)_m=\Gamma(a+m)/\Gamma(a)$ }. Analogously, we have computed the amplitude for a chiral operator of dimension $\Delta$ with two spins \eqref{CSS}, which we can rewrite as
\be
\langle V_C V_{S_2} V_{S_3}\rangle  =\sum_{k=0}^{S_2} (\alphap/2)^{S_2/2+S_3/2-k}  (\epsilon^2\cdot k_3^{S_2-k})(\epsilon^3\cdot k_1^{S_3-k})\eta_{23}^{k} \, \mathcal {C}_k \,,
\ee
where
\be
\mathcal {C}_k =-\frac{\mathcal D_k}{\sqrt 2} +\sum_{l=0}^{k} \frac{n_2! n_3!\,\delta}{4(l-1)!(k-l-1)!(s_2-l)!(s_3-l)!(s_2-k+l)!(s_3-k+l)!} \,,
\ee
and the values of $\delta$ for two massive spins, one massless spin of dimension $\Delta$ and one massive spin, and two spins with $S=2$ are
\be
\delta_{SS}= \frac{\Delta^2}{4\sqrt\lambda}\,, \qquad \delta_{GS}=0 \,,\qquad \delta_{GG}= -\frac{4\alpha_1\alpha_2 \alpha_3\Sigma}{\Delta^2 \sqrt\lambda}\,,
\ee
respectively.
Taking the expression for the structure constant \eqref{structuretwoS} for a chiral primary with scaling dimension $\Delta$ and two massive Regge states with the same spin $S$, in the limit of large coupling constant
we have
\be
C_{\op L SS}^k = \frac{2^{k-S}\Delta^{2S-2k}\lambda^{k/2-S/2+1/4}\Gamma(\Delta/2)^2}{N \sqrt{(\Delta-1)(S-2)}\Gamma(\Delta-1)}  \sum_{l=k}^{S} \binom{l}{k} (2S-4)^{k-l} \mathcal C_l \,.
\ee
If we take one of the Regge spins to have $S=2$ and the opposite $R$-charge of the scalar, then for  $\Delta\ll\lambda^{1/4}$ the structure constant is
\be
C_{\op L GS}^k = \frac{(-1)^{\Delta}\pi^{\frac{3}{2}}(S-2)^{\frac{S+2\Delta-3k+5}{2}}\Delta^{k-3}\lambda^{\frac{2\Delta-k+5}{4}}\csc(\pi\Delta_S/2)}{2^{S+\Delta-\frac{3k+1}{2}+\Delta_S}N\Gamma(\Delta-1)\Gamma(\Delta+3)}  \sum_{l=k}^{2} \binom{l}{k}  \left(\frac{S-2}{2}\right)^{k-l} \mathcal C_l \,.
\ee
Finally, if both operators in the Regge trajectory are at the massless level and have dimension $\Delta$ as the massless scalar then
\be
C_{\op L GG}^k = \frac{8^{k-2}\Delta^{7/2-2k}\lambda^{\frac{k-1}{2}}}{3^{k-1}N} \langle C^\Delta C^\Delta C^\Delta \rangle\binom{2}{k} \left(\frac{4\sqrt\lambda}{3\Delta^2}\right)^{2-k} \mathcal C_{2} \,.
\ee


\subsubsection*{One massive scalar and two Regge spins}
In \appref{sec:contractions} we have computed the two correlators that make up the amplitude for a Konishi with two spins \eqref{V1SS,V2SS}, which we can put together as
\be
\langle V_K V_{S_2} V_{S_3}\rangle  =\sum_{k=0}^{S_2} (\alphap/2)^{S_2/2+S_3/2-k}  (\epsilon^2\cdot k_3^{S_2-k})(\epsilon^3\cdot k_1^{S_3-k})\eta_{23}^{k} \, \mathcal K_k \,.
\ee
The coefficient $\mathcal K_k$ is quite lengthy so we refer the reader to \eqref{coefK}.
Taking the expression for the structure constant \eqref{structuretwoS} for a Konishi and two massive Regge states with the same spin $S$, in the limit of large coupling constant 
\be
C_{K SS}^k = \frac{\sqrt\pi (2\sqrt n-1)^{-\frac{1}{2}+\Delta_S-\Delta_K/2}(2\sqrt n+1)^{-\frac{5}{2}+\Delta_S+\Delta_K/2}}{2^{\Delta_S+\Delta_K-3}(S-2)^{S-k-1+\Delta_S}N}  \lambda^{1/4}\sum_{l=k}^{S} \binom{l}{k} (2S-5)^{k-l} \mathcal K_l \,,
\ee
where $n$ is the mass level of the Regge spins. If we take one of the Regge spins to be at the massless level with scaling dimension $\Delta\ll\lambda^{1/4}$ and the other at the first massive level then the structure constant becomes
\be
C_{KGS}^k= \frac{\Delta^{\frac{9}{2}-k}\lambda^{\frac{k-3}{4}}\Gamma(\Delta/2)^2}{2\sqrt2 N\Gamma(\Delta)}\sum_{l=k}^{2}\binom{l}{k} \left(\frac{\lambda^{1/4}}{\Delta}\right)^{l-k} \mathcal K_l \,.
\ee
Finally, the  structure constant for one Konishi and two states with spin $S=2$ and dimension $\Delta\ll\lambda^{1/4}$ is
\be
C_{KGG}^k= \frac{(-1)^{\Delta}\pi^{\frac{3}{2}}\Delta^{1+2k}\lambda^{\frac{1+2\Delta-2k}{4}}}{2^{k-3+\Delta_K}N \Gamma(\Delta+1)^2} \csc(\pi\Delta_K/2)\sum_{l=k}^{2} \binom{l}{k} (-1)^{l-k} \mathcal K_l \,.
\ee


\section{Discussion}
\label{sec:conclusions}

In the first part of this article we studied how the poles appearing in extremal three-point functions relate 
to the mixing of single- and double-trace operators of \sym. We have made this relation quantitative for the case of the Konishi operator and the double-trace operator made of chiral primaries
with large $R$-charge.

In the paper's second part we computed three-point functions involving higher spin states. For correlators with symmetric traceless operators, conformal symmetry allows for a number of different spacetime dependent structures. Here we have successfully matched the building blocks appearing in the superstring amplitudes with 
the structures allowed by conformal symmetry.  The set of operators we considered are the chiral primaries, the scalar primaries dual to string states at the first massive level, Lagrangian-like operators dual to the dilaton with some Kaluza-Klein momentum and twist two operators dual to strings in the leading Regge trajectory. We have also successfully matched our results for the three-point functions of two massless scalars and a spin state in the leading Regge trajectory with recent calculations using the Mellin amplitude formalism, and older supergravity calculations. Having checked with these known cases we went on to compute new correlation functions involving two higher spin states and Konishi operators.

For the future, there are many interesting directions to explore. For example, it would be interesting to generalize our procedure to include operators with small scaling dimensions. When the operators are dual to strings at massive levels, then our methods are valid,  since 
 their dimensions scale as $2\sqrt n \lambda^{1/4}$. However,  we are forced to give large $R$-charges to protected operators,  which makes it impossible to study three-point functions involving, say,  the chiral primary in the \textbf{20}.  Another direction of study would be to compute sub-leading corrections to our results in the $\ap$ expansion. Besides the corrections coming from the expansions of the dimensions, we would also need to calculate loop corrections to the worldsheet amplitude. 

Another possibility is to study four-point correlators of short operators. In principle the main ideas behind our methods should still hold.  When doing the saddle point approximation one would obtain two intersection points in the bulk, corresponding to the product of two three-point functions and exchange of some operator. It is possible that under some approximations this computation might become feasible.  

Perhaps yet another future  application of our results  involves the comparison to upper bounds of leading twist operators  using  bootstrap methods.   In  \cite{Beem:2013qxa} the authors showed that the upper bounds are consistent with the supergravity correction  in (\ref{djj}) for the $J=2$ double-trace state at large values of the central charge, $c=\sfrac14(N^2-1)$. There is evidence that the upper bounds occur at special values of the coupling  \cite{Beem:2013hha,Alday:2013bha}, which corresponds to an 't Hooft coupling   of order $\la\sim N$.  For this value the splitting is large compared to the supergravity correction but also only affects double-traces with $J\sim N^{1/4}$ and so a present comparison cannot be made.  If instead one could also expand the bootstrap analysis to include the 't Hooft parameter as an input, then one could in principle make such a comparison if at the same time one can find the splitting for low $J$ values. 

Finally, a bigger goal is to understand the structure of the full $\mathrm{AdS}_5\times\mathrm{S}^5$ vertex operators and to  use the underlying integrability of the string world-sheet to compute holographic three-point functions.


\subsection*{Acknowledgments}

We thank T.~Bargheer and V.~Gon\c{c}alves for discussions.
This research is supported in part by
Vetenskapsr{\aa}det under grant \#2012-3269.
JAM thanks the CTP at MIT  for 
hospitality during the course of this work.


\appendix

\addtocontents{toc}{\protect\setcounter{tocdepth}{1}}

\section{Vertex function contractions}
\label{sec:contractions}


\subsection*{Two massless scalars and one Regge spin}
When computing the three-point function of one Regge spin and two massless scalars, we need to compute only two amplitudes. The first, with three NS-NS vertices, is $\langle W_1 W_1 O_{n}\rangle$ which can be computed by using the result for the left movers \eqref{3Regge} and 
taking
$n_1=n_2=0$. 
Combining with the right mover part we have  
\be
\label{W1W1S}
\langle W_1 W_1 O_{n}\rangle  =-\gc^3\left(\sapt\right)^{n-1} \epsilon^1_{M\tilde M}\epsilon^2_{N\tilde N} (\epsilon^3_{RS\tilde R\tilde S} \cdot k_1^{2n-2}) X_L^{MNRS} X_R^{\tilde M\tilde N\tilde R\tilde S} \,,
\ee
with
\be
X_L^{MNRS} = n \eta^{MR}\eta^{NS} + \sapt k_1^R \bigl(k_2^M \eta^{NS} +  k_3^N \eta^{MS} + k_1^S \eta^{MN}\bigr) \,.
\ee
The other correlator is $\langle W_2 W_2 O_{n}\rangle$ which gives
\be
\label{W2W2S}
\langle W_2 W_2 O_n\rangle = - \frac{1}{2}\gc^3 \left(\sapt\right)^n (\epsilon^3_{M\tilde M} \cdot k_1^{2n}) \Tr\left[t^1 \Gamma^M C (t^2)\transpose\Gamma^{\tilde M} C\right] \,.
\ee


\subsection*{One massless scalar, one massive scalar and one Regge spin}
To compute the three-point function of one massless scalar, one scalar at the first massive level and one Regge spin, we will need three amplitudes, two with three NS-NS vertices and one with two vertices in the R-R sector. Once again, for $\langle V_1 W_1 O_{n}\rangle$, we can use the left mover result \eqref{3Regge} and take $n_1=1$ and $n_2=0$. 
Including also the right movers we obtain
\be
\label{V1W1S}
\langle V_{1} W_1 O_{n}\rangle  =  -\gc^3\left(\sapt\right)^{n-2} \epsilon^1_{MN\tilde M\tilde N}\epsilon^2_{P\tilde P} (\epsilon^3_{RST\tilde R\tilde S\tilde T} \cdot k_1^{2n-4}) X_L^{MNPRST} X_R^{\tilde M\tilde N\tilde P\tilde R\tilde S\tilde T}\,,
\ee
with
\begin{align}
 X_L^{MNPRST} =& \left(\sapt\right)^2k_2^M k_1^R k_1^S \left(k_3^P  \eta^{NT}+k_2^N \eta^{PT}+k_1^T \eta^{NP}\right) +n(n-1)\eta^{MR}\eta^{NS}\eta^{PT}\nn\\
&+\sapt k_1^R  \eta^{MS} \left(n k_3^P\eta^{NT} +(1+n) k_1^T \eta^{NP}+2 n k_2^N \eta^{PT}\right) \,.
\end{align}
We also need to compute $\langle V_2 W_1 O_{n}\rangle$ which is
\be
\label{V2W1S}
\langle V_2 W_1 O_n\rangle = -36 \gc^3 \left(\sapt\right)^{n+1}{\alpha^1}^{MNL,\tilde M\tilde N\tilde L} {k_2}_L  {k_2}_{\tilde L}\epsilon^2_{M \tilde M} (\epsilon^3_{N \tilde N} \cdot k_1^{2n})   \,.
\ee
The amplitude with R-R vertices is
\be
\label{V3W2S}
\langle V_3 W_2 O_n\rangle = - \frac{1}{2} \gc^3\left(\sapt\right)^{n-1} (\epsilon^3_{MN\tilde M\tilde N} \cdot k_2^{2n-2}) \Tr\left[t^2 X_L^{MNP} (t_{\tilde P, P}^1)\transpose \left(X_R^{\tilde M\tilde M\tilde P}\right)\transpose\right] \,,
\ee
where
\be
X_L^{MNP} = (n \,\eta^{MP}+\sapt k_3^P k_2^M)(\Gamma^N C)+\sapt k_2^M \eta^{NP}(\slashed k_1 C) \,.
\ee


\subsection*{Two massive scalars and one Regge spin}
In this case we will need three amplitudes with three NS-NS vertices and one with two R-R vertices. For $\langle V_1 V_1 O_{n}\rangle$ we just take \eqref{3Regge} with $n_1 = n_2 =1$, thus obtaining
\be
\label{V1V1S}
\langle V_{1} V_{1} O_{n}\rangle  =-\gc^3\left(\sapt\right)^{n-3} \epsilon^1_{MN\tilde M\tilde N}\epsilon^2_{PQ\tilde P\tilde Q} (\epsilon^3_{RSTV\tilde R\tilde S\tilde T\tilde V} \cdot k_1^{2n-6}) X_L^{MNPQRSTV} X_R^{\tilde M\tilde N\tilde P\tilde Q\tilde R\tilde S\tilde T\tilde V}\,,
\ee
with
\begin{align}
X_L&^{MNPQRSTV} = n\sapt k_1^R\eta^{MS} \eta^{PT}\left(2(n-1) \left( k_3^Q\eta^{NV}+ k_2^N \eta^{QV}\right)+(n+2) k_1^V \eta^{NQ}\right)\nn\\
&+\left(\sapt\right)^2 k_1^R k_1^S\bigg( n \left(k_3^P k_3^Q \eta^{MT}\eta^{NV} + k_2^M k_2^N  \eta^{PT}\eta^{QV}+3 k_2^M k_3^P \eta^{NT}\eta^{QV}\right) \biggr.\nn\\
&\biggl.+ k_1^T \eta^{MP} \left(k_1^V\eta^{NQ}+(n+2) \left(k_3^Q \eta^{NV}+ k_2^N \eta^{QV}\right)\right)\biggr)+n(n-1)(n-2) \eta^{MR}\eta^{NS}\eta^{PT}\eta^{QV} \nn\\
&+\left(\sapt\right)^3 k_2^M k_3^P k_1^R k_1^S k_1^T \left( k_3^Q \eta^{NV}+ k_2^N\eta^{QV}+ k_1^V \eta^{NQ}\right)\,.
\end{align}
The other correlators with NS-NS vertices are
\be
\label{V2V2S}
\langle V_2 V_2 O_n\rangle = -6^2 \gc^3 (\sapt)^{n-1}\alpha^1_{MNL,\tilde M\tilde N\tilde L} \alpha^2_{OPQ,\tilde O\tilde P\tilde Q} \epsilon^3_{RS,\tilde R\tilde S} \cdot k_1^{2n-2}  X_L^{MNLOPQRS} X_R^{\tilde M\tilde N\tilde L\tilde O\tilde P\tilde Q\tilde R\tilde S}
\ee
with
\be
X_L^{MNLOPQRS}=\eta^{NQ}\eta^{LP}\left(\sapt k_1^R\left(\eta^{MO} k_1^S+3(k_3^O \eta^{MS}+k_2^M \eta^{OS})\right)+3n \eta^{MR}\eta^{OS}\right)\,,
\ee
and
\be
\label{V2V1S}
\langle V_2 V_1 O_n \rangle = -36 \gc^3\left(\sapt\right)^{n-1} \alpha^1_{M N L, \tilde M\tilde N\tilde L} \epsilon^2_{PQ,\tilde P\tilde Q} (\epsilon^3_{RS,\tilde R \tilde S} \cdot k_1^{2n-2}) X_L^{MNLPQRS}X_R^{\tilde M\tilde N\tilde L\tilde P\tilde Q\tilde R\tilde S}\,,
\ee
with
\be
X_L^{MNLPQRS} = \sqrt\sapt\eta^{MR} \eta^{NP} k_2^L \left(n \eta^{QS}+\sapt k_3^Q k_1^S \right) \,.
\ee
The correlator with two R-R vertices is
\be
\label{V3V3S}
\langle V_3 V_3 O_n\rangle = - \frac{\gc^3}{2} \left(\sapt\right)^{n-2}(\epsilon^3_{MNL\tilde M\tilde N\tilde L} \cdot k_1^{2n-4}) \Tr\left[t_{\tilde R,R}^1 X_L^{MNLRS} (t_{\tilde S,S}^2)\transpose \left(X_R^{\tilde M\tilde N\tilde L\tilde R\tilde S}\right)\transpose\right] \,,
\ee
where
\begin{align}
&X_L^{MNLRS} = \bigl( n(n-1) \eta^{MR}\eta^{NS}+\sapt k_1^M\left(n (k_3^S\eta^{NR} +\eta^{NS}k_2^R)+k_1^N \eta^{RS}+\sapt k_3^S k_1^N k_2^R\right) \bigr)\left(\Gamma^L C\right) \nn\\
&+\sapt k_1^M \left(\eta^{NS} \bigl(n \eta^{LR}+\sapt k_2^R k_1^L\right) (\slashed k_2 C)-\eta^{NR} \left(n \eta^{LS}+\sapt k_3^S k_1^L\right) (\slashed k_1 C)+\sfrac{\alphap}{4}k_1^{N}\eta^{RS}(\slashed k_1 \Gamma^{L} \slashed k_2 C)\bigr).
\end{align}


\subsection*{One massless scalar and two Regge spins}
When considering the correlator of two Regge spins with a massless scalar it is enough to use \eqref{3Regge} with $n_1=0$ which, for $n_2 \leq n_3$, gives for the left mover contractions
\begin{multline}
\label{W1SSL}
\langle W_{1} O_{n_2} O_{n_3}\rangle_L  =\gc^3 n_2! n_3!\sum_{k_L=0}^{s_2} \frac{(\alphap/2)^{1/2(s_2+s_3-1)-k_L} \eta_{23}^{k_L}}{k_L!(s_2-k_L)!(s_3-k_L)!} \biggl[\sfrac{\ap}{2}k_L(\epsilon^1\cdot k_2)(\epsilon^2 \cdot k_3^{s_2-k_L})(\epsilon^3 \cdot k_1^{s_3-k_L})\biggr. \\
+\biggl. s_2(s_3-k_L){\epsilon^1}^{M}(\epsilon^2\cdot k_3^{s_2-k_L})(\epsilon^3_M\cdot k_1^{s_3-k_L-1})+s_3(s_2-k_L){\epsilon^1}^{M}(\epsilon^2_M\cdot k_3^{s_2-k_L-1})(\epsilon^3\cdot k_1^{s_3-k_L})\biggr]  \,.
\end{multline}
For the right movers we will have the same formula, with the sum running over $k_R$ instead. If we contract with the polarization of the dilaton and denote $k=k_L+k_R$ then we have
\begin{multline}
\label{DSS}
\langle W_{1}^{\mathrm T} O_{S_2} O_{S_3}\rangle  =\gc^3\sum_{k=0}^{S_2} \sum_{l=\max(0,k-\frac{S_2}{2})}^{\min(\frac{S_2}{2},k)}\frac{(n_2! n_3!)^2(\alphap/2)^{S_2/2+S_3/2-k} (\epsilon^2_{M}\cdot k_3^{S_2-k-1})(\epsilon^3_{N}\cdot k_1^{S_3-k-1})}{l!(s_2-l)!(s_3-l)!(k-l)!(s_2-k+l)!(s_3-k+l)!}\\
\times\eta_{23}^{k}\biggl[ \biggl(\frac{k S_2 S_3}{4}-2 l(k-l)\biggr)k_3^M k_1^N+\frac{S_2 S_3}{4\ap}\biggl(k(S_2+S_3)-S_2 S_3-4 l(k-l)\biggr) \eta^{MN}\biggr] \,.
\end{multline}
If instead we contract with the polarization of the chiral operator then we obtain
\begin{multline}
\label{CSS}
\langle W_{1} O_{S_2} O_{S_3}\rangle  =\gc^3\sum_{k=0}^{S_2} \sum_{l=\max(0,k-\frac{S_2}{2})}^{\min(\frac{S_2}{2},k)}\frac{(n_2! n_3!)^2(\alphap/2)^{S_2/2+S_3/2-k} (\epsilon^2_{M}\cdot k_3^{S_2-k-1})(\epsilon^3_{N}\cdot k_1^{S_3-k-1})}{l!(s_2-l)!(s_3-l)!(k-l)!(s_2-k+l)!(s_3-k+l)!}\\
\times\eta_{23}^{k}\biggl[ \biggl(\frac{-k S_2 S_3}{4}+(2+\delta) l(k-l)\biggr)k_3^M k_1^N-\frac{S_2 S_3}{4\ap}\biggl(k(S_2+S_3)-S_2 S_3-4 l(k-l)\biggr)\eta^{MN}\biggr] \,,
\end{multline}
with
\be
\delta = \frac{1}{\sqrt\lambda\Delta_1^2}\left(\sqrt\lambda((S_2+S_3-4)\Delta_1^2-(S_2-S_3)(\Delta_2^2-\Delta_3^2))+\lambda (S_2-S_3)^2-4\alpha_1\alpha_2\alpha_3\Sigma\right)\,.
\ee


\subsection*{One massive scalar and two Regge spins}
For this correlator there are two amplitudes we must consider. The first one can be taken by using \eqref{3Regge} with $n_1=1$ which, for $n_2 \leq n_3$, gives for the left mover contractions
\begin{multline}
\label{V1SSL}
\langle V_{1} O_{n_2} O_{n_3}\rangle_L  =\gc^3 n_2! n_3!\sum_{k_L=0}^{s_2} \frac{(\alphap/2)^{1/2(s_2+s_3)-k_L-1} \eta_{23}^{k_L}}{k_L!(s_2-k_L)!(s_3-k_L)!}\epsilon^1_{MN}(\epsilon^2_{PQ} \cdot k_3^{s_2-k_L-2})(\epsilon^3_{RS} \cdot k_1^{s_3-k_L-2}) \\
\times\biggl[s_3 (s_2-k_L)(n_2-k_L) \eta^{MP}\eta^{NQ} k_1^R k_1^S+s_2 (s_3-k_L)(n_3-k_L) \eta^{MR}\eta^{NS} k_3^P k_3^Q\biggr. \\
+(s_2+s_3-1)(s_2-k_L)(s_3-k_L) \eta^{MP}\eta^{NR} k_3^Q k_1^S+(\sfrac{\ap}{2})^2 k_L k_2^M k_2^N k_3^P k_3^Q k_1^R k_1^S\\
\biggl. +\sfrac{\ap}{2} k_2^M k_3^P  k_1^R \bigl((s_3+k_L)(s_2-k_L) \eta^{NQ} k_1^S +(s_2+k_L)(s_3-k_L) \eta^{NS} k_3^Q \bigr)\biggr]\,.
\end{multline}
For the right movers we will have an analogous formula, so contracting with the polarization of the Konishi opeartor and using $k=k_L+k_R$ leads to
\begin{multline}
\label{V1SS}
\langle V_{1} O_{n_2} O_{n_3}\rangle=\gc^3\sum_{k=0}^{S_2}\sum_{l=\max(0,k-\frac{S_2}{2})}^{\min(\frac{S_2}{2},k)} \frac{(n_2! n_3!)^2(\alphap/2)^{\frac{S_2+S_3}{2}-k} (\epsilon^2_{MN}\cdot k_3^{S_2-k-2})(\epsilon^3_{PQ}\cdot k_1^{S_3-k-2})\eta_{23}^{k}}{l!(s_2-l)!(s_3-l)!(k-l)!(s_2-k+l)!(s_3-k+l)!}\\
\times\biggl[ K_1(k,l,S_i,\Delta_i) k_3^M k_3^N k_1^P k_1^Q+\sfrac{2}{\ap}K_2(k,l,S_i,\Delta_i)\eta^{MP} k_3^N k_1^Q+\sfrac{4}{\ap^2} K_3(k,l,S_i,\Delta_i)\eta^{MP} \eta^{NQ}\biggr] \,,
\end{multline}
with 
\begin{align}
&K_1(k,l,S_i,\Delta_i)=\frac{k S}{32\lambda}(\Delta_-^2-4\sqrt\lambda(\Delta_++S_-\Delta_-))-\frac{l(k-l)}{256\lambda^{2}}\Delta_-^2(\Delta_-^2-8\sqrt\lambda(2\Delta_++S_-\Delta_-))\nn\\
&+\frac{S}{96}(4S_+ k(3 S_+ +2)-(3S+28k+48 S k-12k^2))-\frac{ l(k-l)}{288\lambda}\biggl(144\lambda l(k-l)\biggr.  \nn\\
&\qquad\qquad-36\sqrt\lambda((S_-(S_-^2+2S_+ +4k)+4S_-)\Delta_-+2(2k-6+2S_+ +S_-^2)\Delta_+)\nn\\
&\qquad\qquad+9((3S_-^2+2(S_+ +2k+1))\Delta_-^2+8 S_-\Delta_-\Delta_+-16\Delta_2^2\Delta_3^2)+2\lambda\biggl(9S_+^4+36S_+^3\bigr.\nn\\
&\qquad\qquad\biggl.-2(11+36(S-k))S_+^2-8(13+18S-15k)S_+ +4(56+9S(3+4S-8k))\biggr)\biggr)\,,\nn\\
&K_2(k,l,S_i,\Delta_i)=\frac{l(k-l)}{144}\biggl(9S_+^4+22S_+^3+2S_+^2(13k-45S-32)+4S_+(10-27S-8k)\biggr.\nn\\
&\qquad\qquad\qquad\qquad\qquad\qquad\biggl.+8(27S(S+1)-5k)-12l(k-l)(3S_-^2+16(S_+-2))\biggr)\nn\\
&-\frac{S_2S_3}{576}\bigl(S_+^2(9S+34k-36k^2)-4S_+(3S+22k+18k^2)+8(3S-4S^2+5k+18k^2(1+S))\bigr)\nn\\
&-\frac{1}{256\lambda}(S(S-4k^2)+4l(k-l)(4l(k-l)-(S_-^2-2S)))\bigl(\Delta_-^2-4\sqrt\lambda(2\Delta_++S_-\Delta_-)\bigr)\,,\nn\\
&K_3(k,l,S_i,\Delta_i)=\frac{S}{1152}\biggl(4S_+(25S+50k+18kS+64k^2) -S_+^2(7S+128k-100k^2)\biggr.\nn\\
&\qquad\qquad\qquad\qquad\qquad\qquad\biggl.+14kS_+^3-2(18S^2+86S+72Sk+57k^2-72Sk^2)\biggr)\nn\\
&\;+\frac{l(k-l)}{288}\biggl(-7S_+^4+14(k+2)S_+^3+(50S_2S_3-56k-28)S_+^2+8(S(9k+2)+7k)S_+\biggr.\nn\\
&\qquad\qquad\qquad\biggl.-8(27S^2+18kS+11)-4l(k-l)(7(S_+-2)^2+36S)\biggr)\,,
\end{align}
where we defined $S_-=S_2-S_3$, $S_+=S_2+S_3$, $S=S_2 S_3$, $\Delta_-=\Delta_2^2-\Delta_3^2$ and $\Delta_+=\Delta_2^2+\Delta_3^2$.
The other correlator needed is, for $n_2 \leq n_3$
\be
\label{V2SSL}
\langle V_2 O_{n_2} O_{n_3} \rangle_L = \gc^3 n_2! n_3!\sum_{k_L=0}^{n2} \frac{6\left(\alphap/2\right)^{-k_L+\frac{n_2+n_3+1}{2}}\eta_{23}^{k_L}}{k_L! (n_2-k_L)!(n_3-k_L)!}\alpha_1^{MNL}\epsilon^2_{M}\cdot k_3^{n_2-k_L}\epsilon^3_{N}\cdot k_1^{n_3-k_L}  {k_3}_L .
\ee
Combining with the contractions from the right movers and contracting indices from the scalar polarization, we get for $k=k_L+k_R$ 
\begin{multline}
\label{V2SS}
\langle V_2 O_{S_2} O_{S_3}\rangle  = \gc^3\sum_{k=0}^{S_2}\sum_{l=\max(0,k-\frac{S_2}{2})}^{\min(\frac{S_2}{2},k)} \frac{(n_2! n_3!)^2(\alphap/2)^{\frac{S_2+S_3}{2}-k} (\epsilon^2_{MN}\cdot k_3^{S_2-k-2})(\epsilon^3_{PQ}\cdot k_1^{S_3-k-2})\eta_{23}^{k}}{l!(n_2-l)!(n_3-l)!(k-l)!(n_2-k+l)!(n_3-k+l)!}\\
\times\biggl[ -\frac{1}{2} k_3^M k_3^N k_1^P k_1^Q+\frac{2}{\ap}K_4(S_i,\Delta_i)\eta^{MP} k_3^N k_1^Q+\frac{4}{\ap^2} K_5(S_i,\Delta_i)\eta^{MP} \eta^{NQ}\biggr] \,,
\end{multline}
where 
\begin{align}
K_4(S_i,\Delta_i)&=-\frac{1}{16\lambda}( 4\lambda(S_-^2-2S_+ +4)+\Delta_-^2+4\sqrt\lambda (2\Delta_+-S_-\Delta_-))\,,\nn\\
K_5(S_i,\Delta_i)&=\frac{1}{16\lambda}(2(S_-^2-4S_++12)\lambda-\Delta_-^2+8\sqrt\lambda \Delta_+)\,.
\end{align}
Putting the two correlators together we get
\be
\langle V_K V_{S_2} V_{S_3}\rangle  =\sum_{k=0}^{S_2} (\alphap/2)^{S_2/2+S_3/2-k}  (\epsilon^2\cdot k_3^{S_2-k})(\epsilon^3\cdot k_1^{S_3-k})\eta_{23}^{k} \, \mathcal K_k \,,
\ee
with the coefficients $\mathcal K_k$ given by
\begin{align}
\label{coefK}
\mathcal K_k=& \gc^3 \sum_{l=\max(0,k-\frac{S_2}{2})}^{\min(\frac{S_2}{2},k)} \frac{n_2!n_3!}{16l!(s_2-l)!(s_3-l)!(k-l)!(s_2-k+l)!(s_3-k+l)!}\times\nn\\
&\times\biggl[K_1(k,l,S_i,\Delta_i)+\frac{(k-l)}{(s_2-k+l+1)(s_3-k+l+1)}K_2(k-1,l,S_i,\Delta_i)\biggr.\nn\\
&\quad+\frac{(k-l)(k-l-1)}{(s_2-k+l+1)(s_3-k+l+1)(s_2-k+l+2)(s_3-k+l+2}K_3(k-2,l,S_i,\Delta_i)\nn\\
&\quad-\frac{(s_2-k+l)(s_3-k+l)(s_2-l)(s_3-l)}{2}+(k-l)(s_2-l)(s_3-l)K_4(S_i,\Delta_i)\nn\\
&\quad\biggl.+\frac{(k-l)(k-l-1)(s_2-l)(s_3-l)}{(s_2-k+l+1)(s_3-k+l+1)}K_5(S_i,\Delta_i)\biggr]\,.
\end{align}

\providecommand{\nbbststyle}{\raggedright\footnotesize\parskip0pt}
\bibliographystyle{nb}
\bibliography{references_jhep}

\end{document}

%% file: drawing.pdf_tex
\begingroup%
  \makeatletter%
  \providecommand\color[2][]{%
    \errmessage{(Inkscape) Color is used for the text in Inkscape, but the package 'color.sty' is not loaded}%
    \renewcommand\color[2][]{}%
  }%
  \providecommand\transparent[1]{%
    \errmessage{(Inkscape) Transparency is used (non-zero) for the text in Inkscape, but the package 'transparent.sty' is not loaded}%
    \renewcommand\transparent[1]{}%
  }%
  \providecommand\rotatebox[2]{#2}%
  \ifx\svgwidth\undefined%
    \setlength{\unitlength}{591.77912598bp}%
    \ifx\svgscale\undefined%
      \relax%
    \else%
      \setlength{\unitlength}{\unitlength * \real{\svgscale}}%
    \fi%
  \else%
    \setlength{\unitlength}{\svgwidth}%
  \fi%
  \global\let\svgwidth\undefined%
  \global\let\svgscale\undefined%
  \makeatother%
  \begin{picture}(1,0.63413452)%
    \put(0,0){\includegraphics[width=\unitlength]{drawing.pdf}}%
    \put(0.25273567,0.50066279){\color[rgb]{0,0,0}\makebox(0,0)[lb]{\smash{$\epsilon_1$}}}%
    \put(0.7170489,0.50589698){\color[rgb]{0,0,0}\makebox(0,0)[lb]{\smash{$\epsilon_3$}}}%
    \put(0.50007766,0.15629883){\color[rgb]{0,0,0}\makebox(0,0)[lb]{\smash{$\epsilon_2$}}}%
    \put(0.23855662,0.58987767){\color[rgb]{0,0,0}\makebox(0,0)[lb]{\smash{$k_2$}}}%
    \put(0.16906039,0.45861904){\color[rgb]{0,0,0}\makebox(0,0)[lb]{\smash{$k_2$}}}%
    \put(0.73107943,0.59180298){\color[rgb]{0,0,0}\makebox(0,0)[lb]{\smash{$k_1$}}}%
    \put(0.8005993,0.4625524){\color[rgb]{0,0,0}\makebox(0,0)[lb]{\smash{$k_1$}}}%
    \put(0.57022374,0.09179896){\color[rgb]{0,0,0}\makebox(0,0)[lb]{\smash{$k_3$}}}%
    \put(0.43438399,0.09681932){\color[rgb]{0,0,0}\makebox(0,0)[lb]{\smash{$k_3$}}}%
    \put(0.87721084,0.5548475){\color[rgb]{0,0,0}\makebox(0,0)[lb]{\smash{$S_3-j-k$}}}%
    \put(0.44709765,0.00637643){\color[rgb]{0,0,0}\makebox(0,0)[lb]{\smash{$S_2-i-k$}}}%
    \put(-0.00224429,0.56159617){\color[rgb]{0,0,0}\makebox(0,0)[lb]{\smash{$S_1-i-j$}}}%
    \put(0.3527625,0.35014466){\color[rgb]{0,0,0}\makebox(0,0)[lb]{\smash{$i$}}}%
    \put(0.60078761,0.29062027){\color[rgb]{0,0,0}\makebox(0,0)[lb]{\smash{$k$}}}%
    \put(0.53530042,0.52532908){\color[rgb]{0,0,0}\makebox(0,0)[lb]{\smash{$j$}}}%
  \end{picture}%
\endgroup%